\documentclass[12pt]{article}
\usepackage{leqno,amsmath,amssymb,graphicx,color,ulem,epstopdf}

\newcommand{\mbf}[1]{\mbox{\boldmath $#1$}}
\newcommand{\ba}{{\mbf \beta}}

{\catcode `\@=11 \global\let\AddToReset=\@addtoreset}
\AddToReset{equation}{section}

\AddToReset{Theorem}{section}

\newtheorem{rema}{Remark}[section]
\newtheorem{theo}{Theorem}[section]
\newtheorem{prop}{Proposition}[section]

\newcommand{\cH}{{\cal H}}

\def\ba{\begin{array}}
\def\bc{\begin{center}}
\def\bd{\begin{description}}
\def\be{\begin{enumerate}}
\def\ea{\end{array}}
\def\ec{\end{center}}
\def\ed{\end{description}}
\def\edt{\end{document}}
\def\ee{\end{enumerate}}
\def\ben{\begin{equation}}
\def\benn{\begin{equation*}}
\def\een{\end{equation}}
\def\eenn{\end{equation*}}
\def\benr{\begin{eqnarray}}
\def\eenr{\end{eqnarray}}
\def\benrr{\begin{eqnarray*}}
\def\eenrr{\end{eqnarray*}}

\def\al{\alpha}
\def\b{\beta}

\def\D{\Delta}

\def\edt{\end{document}}
\def\ep{\epsilon}
\def\g{\gamma}

\def\h{\hat}

\def\hs{\hskip}
\def\iny{\infty}

\def\L{\Lambda}
\def\la{\lambda}
\def\lel{\label}

\def\m{\mu}
\def\mb{\mbox}
\def\noi{\noindent}

\def\nn{\nonumber}

\def\r{\ref}

\def\si{\sigma}

\def\sti{\sum_{i=1}^n}
\def\stj{\sum_{j=1}^n}

\def\Si{\Sigma}
\def\t{\tau}

\def\ti{\tilde}

\def\vep{\varepsilon}

\def\vs{\vskip}

\def\R{{\mathbb R}}

\def\z{\zeta}

\def\bX{{\mbf X}}

\begin{document}

\def\tecr{\textcolor{red}}
\def\tecb{\textcolor{blue}}
\def\so{\sout}
\def\hDash{\bot\!\!\!\bot}
\def\Cov{\mb{Cov}}
\def\Var{\mb{Var}}

\newtheorem{step}{\sc Step}[section]

\bc
{\large \bf An Adaptive-to-Model Test for Parametric Single-Index Errors-in-Variables Models}\\[.3cm]
Hira L. Koul, \,\,\, Chuanlong Xie, \,\,\,  Lixing Zhu\footnote{The corresponding author. Email: lzhu@hkbu.edu.hk. The research described here was supported by a grant from the Research Council of Hong Kong, and a grant from Hong Kong Baptist University, Hong Kong. This is a part of the PHD thesis of the second author.}\\
\small\it Michigan State University, USA\\
\small\it  Hong Kong Baptist University, Hong Kong, China

\ec

\date{}

\begin{abstract}
This paper provides some useful tests for fitting a parametric single-index regression model when covariates are measured with error and validation data is available.
We propose  two tests whose consistency rates do not depend on the dimension of the covariate vector when an adaptive-to-model strategy is applied.
One of these tests has  a bias term that becomes arbitrarily large with increasing sample size but its asymptotic variance is smaller, and the other is asymptotically unbiased with larger asymptotic variance.
Compared with the existing local smoothing tests, the new tests behave like a classical local smoothing  test with only one covariate, and still are omnibus against general alternatives.
This avoids the difficulty associated with the curse of dimensionality.
Further, a systematic study is conducted to give an insight on the effect of the values of the ratio between the sample size and the size of validation data on the asymptotic behavior of these tests.
Simulations are conducted to examine the performance in several finite sample scenarios.
\end{abstract}

\section{Introduction}

Consider the nonparametric regression model with measurement error where the response variable $Y$, a $p$-dimensional unobservable predicting covariate $X$ and its observable cohort vector $W$ are related to each other by the relations
\benr\label{1}
    Y=\m(X) +\vep, \quad
    W=X+U.
\eenr
Here $p$ is assumed to be known, and the variables $\vep,\, U,$ and $X$ are assumed to be mutually independent with $E(\vep)=0=E(U)$.
Hence $\m(x)=E(Y|X=x)$ is the usual regression function.
This is the so called nonparametric errors in variables (EIVs) regression model.
The monographs of Fuller (1987), Cheng and Van Ness (1999), and Carroll, Ruppert, Stefansky and Crainiceanu (2006) contain a vast number of real data examples where this model is naturally applicable.

The problem of interest here is to fit a parametric single-index  regression model to the regression function,
i.e., for a known real valued link function $g$ we wish to test the hypothesis
\benrr
&& H_0:\,\, \m(x)=g(\b^{\top} x), \quad \mb{for all $x\in \R^p$ and for some $\b\in \R^p$},\,\,\,\mb{versus}\\
&& H_1:\,\,\mb{$H_0$ is not true.}
\eenrr
Throughout this paper,  $a^{\top}$ denotes transpose of the vector $a\in \R^p$. The model is called parametric single index although it is also often called generalized linear model. This is because it is in effect slightly different from the generalized linear model that has its special definition in the literature.
A motivation for considering the above testing problem is that in practice model checking is necessary to prevent possible wrong conclusions when an improper model is used.
Moreover, efficient and accurate inference is possible in a parametric model than in a nonparametric or semiparametric model.

Hart (1997) described numerous tests for lack-of-fit of a parametric regression model in the classical regression set up where $X$ is observable. Since the mid 1990's, there has been an explosion of activities in this area as is summarized in the recent review by González-Manteiga and  Crujeiras (2013).

It is well known that the naive application of the inference procedures valid for the classical regression set up,  where one replaces $X$ by $W$, often yields inefficient inference procedures for the EIV models, see, e.g. Fuller (1987) and Carroll et al. (2006).
An alternative approach adopted in the literature is that of  calibration, where the original regression relationship is transferred to the regression $E(Y|W)$ relationship between the response $Y$ and the cohort $W$.
Zhu, Cui and Ng (2004) established a sufficient and necessary condition for the linearity of $E[Y|W]$ with respect to $W$ when $g(\b^{\top} x)=\al+\b^{\top} x$.
A score-type lack-of-fitness test was proposed based on this fact.
This testing procedure has been extended to polynomial EIVs models by Cheng and Kukush (2004) and Zhu, Song and Cui (2003) independently, without the normality restriction on the covariates.
Hall and Ma (2007) proposed a test based on deconvolution methods assuming that the distribution of the covariate errors is known.
Zhu and Cui (2005) proposed a test for fitting a general linear model $\al+\b^{\top}h(x)$, where $h$ is a vector of known functions.
Song (2008) proposed a test for fitting $\b^{\top} h(x)$ to $\m(x)$, without requiring the knowledge of the density of $X$.
He used the deconvolution kernel density estimator.
Koul and Song (2009) developed an analog of the minimum distance tests of Koul and Ni (2004) to fit a parametric form to the regression function for the Berkson measurement error models.
Koul and Song (2010) developed tests for fitting a parametric function to the nonparametric part in a partial linear regression model under a similar condition.
These latter five references assume that density of the measurement error $U$ is known.
All of these authors employ the calibrated methodology and test for fitting the parameter form of the regression function $E[Y|W]$ implied by $H_0$.

There is no valid test in the literature for fitting a parametric model under general conditions where the distributions of both $X$ and $U$ may not be known.
Some of the main reasons for this are the difficulties associated with the estimation of the calibrated regression function and some of the other underlying functions involved in the construction of a test statistic.
However, it is possible to circumvent some of these difficulties  when there are validation data available.
Stute, Xue and Zhu (2007) used validation data and empirical likelihood methodology to develop confidence regions for some underlying parameters. Song (2009) developed a test for general EIVs models with the assistance of validation data without assuming any knowledge of the distributions of $X$ or $U$, under somewhat restrictive conditions on the kernel function and bandwidth.
Dai, Sun and Wang (2010) constructed a test with validation data for the same model as in Zhu and Cui (2005).
They used specific models and relaxed some conditions in Song (2009).
Xu and Zhu (2014) considered a nonparametric test for partial linear EIVs models with validation data.
All of these tests are based on local smoothing methodology.

In the classical regression setup, it is known that a common property of lack-of-fit tests
for fitting a parametric regression model based on nonparametric smoothing methodology is
that the rate of consistency of the test statistics is $1/\sqrt {nh^{p/2}}$.
That is, the null distribution of a suitably centered and scaled test statistic multiplied by
$\sqrt {nh^{p/2}}$ has a weak limit, and these
tests can detect local alternatives distinct from the null only at this rate.
When $p$ is even 2 or larger, this rate can be very slow.
Consequently, for moderate sample sizes, local smoothing tests cannot maintain the
significance level well and have low power even for $p=2$ or $3$. See, e.g.,
 Zheng (1996), Koul and Ni (2002), and several other cited references for this phenomena.
It is expected that the same fact will continue to hold for various local smoothing
tests in the EIVs setup.

The main goal of the present paper is to propose  tests of dimension reduction nature
when validation data is available, which do not suffer from the above slow rate
of consistency. Specifically, the tests do not suffer severely from the curse of
dimensionality and can well maintain the significance level with good power performance
for moderate finite sample sizes.
Towards this goal we proceed as follows.
First, we discuss sufficient dimension reduction (SDR) technique as illustrated in Cook
(1998), Li and Yin (2007), and Carroll and Li (1992).
The goal is to have a technique such that the dimension of $X$ can be reduced to
one-dimensional projection $\b^{\top}X$ under the null hypothesis, where $\b$ is just the
projection direction in the model (\ref{1}) and to $B^{\top}X$ automatically under the
alternative, where $B$ is a $p\times q$ orthonormal matrix with $q\le p$ to be specified.
Second, based on  dimension reduction, we can then construct a test with the consistency
rate of $1/\sqrt{nh^{1/2}}$ (or $1/(nh^{1/2})$ when a quadratic form is used) when the
size $N$ of  validation data is proportional to or larger than the sample size $n$.
When $N$ is much smaller than $n$, the consistency rate can be slower.
Therefore, the third issue is to investigate the relationship between the asymptotic
behaviour of the tests and the size of validation data set.
In Section~3, a systematic study is performed to analyze the three different scenarios:
$N/n\to \la$, as $\min(n, N)\to \iny$,
where $\la=0,\, \iny,$ or $0<\la<\iny$.
Another interesting issue is raised during the construction procedure.
When validation data are used to define the nonparametric kernel estimate of $E(Y|W)$ such
that the residuals can be derived, the resulting test would have a bias term going to
infinity as $n \to \iny$. It motivates us to consider a bias correction.

To efficiently employ sufficient dimension reduction theory (SDR) of Cook (1998) or CMS of
Cook and Li (2002), we consider the alternatives $\ti H_1: \m(x)=G(B^{\top}x),$  for all $x\in
\R^p$, and for some $p\times q$ orthonormal matrix $B$ with an unknown $q\le p$ and for
some real valued function $G$.
When there are no measurement errors in covariates, Guo, Wang and Zhu (2015) proposed a
dimension-reduction  model-adaptive approach to circumvent the dimensionality problem.
To implement this methodology one needs to estimate  the matrix $B$.
There are a number of proposals available in the literature for this purpose.
Examples include sliced inverse regression (SIR) of Li (1991), sliced average variance
estimation (SAVE) of Cook and Weisberg (1991),
contour regression (CR) of Li et al. (2005), directional regression (DR) of Li and Wang
(2007), discretization-expectation estimation (DEE) of Zhu et al. (2010a), and the average
partial mean estimation (APME) of  Zhu et al. (2010b).

In this paper,  we  construct an adaptive-to-model test in the current set up.
The proposed test is based on the Zheng's test (1996).
To this end, we consider a different kind of calibration where instead of conditioning
on $W$ we condition on $\b^{\top}W$ under the null hypothesis and
on $B^{\top}W$ under the alternatives, and then constructs a test for this
testing problem. Thus, our strategy is sketched as follows: 1). Use the data $(w_1, y_1),
\cdots, (w_n, y_n)$ to estimate $\b$ under the null hypothesis and automatically the
matrix $B$ by a $q\times q$ orthogonal matrix $C$ under the alternative;
2). Use the validation data to estimate the conditional expectation
$E[g(\b^{\top}X)|\b^{\top}W]$.
3). Compute the test statistic using these regression function estimates.

As mentioned above, the test statistic is asymptotically biased.
It is because of the dependence among the residuals when we use all the validation data to
obtain the estimators in Step 2.
To reduce the bias, we propose a bias correction method to construct another test.
In the simulation studies, we will compare their performance.

The paper is organized as follows:  Section~2 contains  a brief description of the test
statistic construction. Since the estimation the matrix $B$ plays a key role in
having the dimension reduction property of the test, we review a widely used
dimension reduction method in this section. The needed assumptions are also stated in this
section. The asymptotic properties of the test
statistic under the null and alternative hypotheses are described in Section 3.
Particularly, a systematic study is conducted on the asymptotic behaviors of the tests
under the three scenarios where the ratio $N/n$ of the validation data $N$ and the  sample size $n$ is small, moderate and large. Section 4 presents the simulation
results. The proofs are postponed to Appendix.

Before closing this section, we describe some notation used in the sequel.
The sample is denoted by $\{(y_{i}, w_{i}), i=1, \cdots, n\}$  and the validation data is dented
by $\{(\ti w_{s}, \ti x_{s}),s=1,\cdots,N\}$. The two data sets are assumed to independent of each other.
Further, in various expressions below, $i$ and $j$ often represent the indices of primary data, while $s$ and $t$ those of validation data.
Throughout this paper, $\to_p$ denotes the convergence in
probability and "$\to_D$" stands for the convergence in distribution.
All limits are taken as $n\wedge N\to \iny$, unless specified otherwise. The normal distribution with
mean $a$ and variance $b$ is denoted by $N(a,b)$.

\section{Methodology development}

\subsection{Test construction: a dimension-reduction adaptive-to-model strategy }
In this subsection, we describe the details of test statistics construction.
It consists of three components as follows.\\
1). {\it Model adaptation}.
To proceed further, let $ r( w,\b)=E[g(\b^{\top}X)|W=w]$,  $w\in \R^p$, denote the
new regression function under the null hypothesis.
In order to avoid the above mentioned high dimensionality problem of nonparametric
estimators of $r(\cdot, \cdot)$  due to the dimension of $W$, we adopt the following
dimension reduction adaptive-to-model strategy (DREAM).
Recall that $W=X+U$. Note that under $H_0$, the regression function $g(\b^{\top}X)$
depends on $X$ only through the linear combination $\b^{\top}X$. It is then natural to
consider the situation where the calibrated regression function $E(Y|W)$ depends on
$W$ only through a linear combination of the
components of $W$, i.e., when $E(Y|W)=E[g(\b^{\top}X)|W]=E[g(\b^{\top}X)|\b^{\top}W]:=r(\b^{\top}W, \b)$. Similarly, under
the alternative, we assume that $E(Y|W)=E(Y|B^{\top}W)=E(G(B^{\top}X)|B^{\top}W)$.
Thus the transferred hypotheses become as follows:
\benr\label{5}
\cH_{0}: P\{E(Y|W)=r(\b^{\top}W,\b)\}=1, \quad \text{ for some $\b\in \R^p$},
\eenr
versus the transferred alternative hypothesis:
\benr\label{6}
\cH_{1}: P\{E(Y|W)= E(Y|B^{\top}W)\not =r(\b^{\top}W,\b)\}=1, \quad \text{for all $\b\in \R^p$ }.
\eenr
Generally the two hypotheses $H_0$ and $\cH_0$ are not exactly equivalent.
But, as in Song (2008), when the family densities $f_{\b^{\top}U}(\b^{\top}w - \cdot)$ is a complete family over the parameter $\b^{\top}w\in \R$, the equivalence can hold.

2). {\it Test statistic construction.} Let $e=Y-r(\b^{\top}W, \b)$.
To unify the null and alternatives, let $B=\b c$ under $\cH_{0}$ where $c$ is a constant, hence $E[e|\b^{\top}W]=E[e|B^{\top}W]=0$.
Moreover, following Zheng (1996),
\benrr
    E[e E[e|\b^{\top}W]f(\b^{\top}W)]=E[E^2(e|\b^{\top}W)f(\b^{\top}W)]
    =E[E^2(e|B^{\top}W)f(B^{\top}W)]=0,
\eenrr
and under $\cH_1$, $E[E^2(e|B^{\top}W)f(B^{\top}W)]>0$.
To obtain residuals for the construction of the test statistics, we assume the availability of validation data $(w_s,x_s),s=1,\cdots,N$, which is used to estimate the function $r$.
Note that $r$ is an unknown function of $\b^{\top}W$.
In order to construct an estimator $r(\b^{\top}W, \b)$, let $M(\cdot)$ be a kernel function, $v_N$ be a bandwidth sequence, and set
$M_{v_N}(\cdot)=v_N^{-1}M(\cdot/v_N)$.
Then an estimator of $r(\b^{\top}W, \b)$ is
\benr\label{rhat}
\h r(\h \b^{\top}w,\h \b)=\frac{\sum_{s=1}^{N}M_{v_{N}}(\h \b^{\top} w-\h \b^{\top} \ti w_s)
 g(\h \b^{\top} \ti x_s)}{\sum_{s=1}^{N}M_{v_{N}}(\h \b^{\top}w-\h \b^{\top} \ti w_s)},
\eenr
where  $\h \b$ is a consistent estimate of  $\b$ based on primary data. Define the residuals
\benr\lel{res}
e_i=y_i-r(\b^{\top}w_i,\b), \quad \h e_i=y_i-\h r(\h \b^{\top}w_i,\h \b), \quad i=1,\cdots, n.
\eenr

To estimate the conditional expectation of the error $e$, given $B^{\top}W$, we also need an
estimator $\h B(\h q)$  of $B$ that is consistent to $\b/\|\b\|$ under the null, and to $B$
under the alternative. This model adaptation property of $\h B(\h q)$  can enable the test
statistic to adapt to model and then to alleviate the curse of dimensionality. This estimator
will be specified later. For the moment assume the existence of such an estimator.

To proceed further, let $K$ be another kernel function and $h\equiv h_n$
another bandwidth. Then an estimator of the product $E[e|B^{\top}W]f(B^{\top}W)$ at
$\h B^{\top}w_{i}$ is given by
\benrr
\h E[e_i|\h B(\h q)^{\top}w_i]\h f(\h B(\h q)^{\top}w_i)&=&
\frac{1}{n-1}\sum_{j\neq i}^{n} K_h(\h B(\h q)^{\top}w_j
- \h B(\h q)^{\top}w_i)\h e_j.
\eenrr
The analog of the Zheng's test statistic in the current set up is based on an estimator
of $E[eE[e|W]f(W)]$, given by
\benr\label{7}
 \ti V_n=\frac{1}{n(n-1)}\sum_{i=1}^{n}\sum_{j\neq
 i}^{n}\h e_i K_h(\h B(\h q)^{\top}(w_i-w_j))\h e_j.
\eenr\\
3). {\it Bias correction}.
From the technical details in Appendix, we can see that the test statistic in (\ref{7}) has
non-negligible asymptotic bias and thus its limiting null distribution has a mean tending  to
infinity unless $n/(Nh^{1/2})\to  0$, which makes the bias term vanish.
The main reason is the dependence between the residuals $\h e_i$ and $\h e_j$ for $i\not = j$
when all validation data are used to estimate the function $r$.
There are two ways to correct for this bias. One is to center  the test statistic at a suitable estimator of this bias.
This is a traditional method, and has been used.
Alternately, we propose a block-wise estimation approach to asymptotically eliminate the bias as follows.
Assume $N$ is a positive even integer.
We halve the whole validation data set, use the two halves to construct two estimators of
the regression function $r$, which results in the two sets of residuals as follows. Let
\benr\label{r12hat}
&&\h r_{(1)}(\h \b^{\top}w,\h \b)=\frac{\sum_{s=1}^{N/2}M_{v_{N}}(\h \b^{\top} w-\h \b^{\top}
\ti w_s) g(\h \b^{\top} \ti x_s)}{\sum_{s=1}^{N/2}M_{v_{N}}(\h \b^{\top}w-\h \b^{\top} \ti w_s)},
 \\
&& \h r_{(2)}(\h \b^{\top}w,\h \b)=\frac{\sum_{s=N/2+1}^{N}M_{v_{N}}(\h \b^{\top} w-\h \b^{\top}
\ti w_s)
 g(\h \b^{\top} \ti x_s)}{\sum_{s=N/2+1}^N M_{v_{N}}(\h \b^{\top}w-\h \b^{\top} \ti w_s)}, \nn \\
&& \h e_{i(1)}:= y_i -\h r_{(1)}(\h \b^{\top}w_i,\h \b), \quad \h e_{i(2)}= y_i -\h r_{(2)}(\h
\b^{\top}w_i,\h \b), \quad i=1, \cdots, n. \nn
\eenr

Use these residuals to define the test statistic
\benr\label{8}
V_n=\frac{1}{n(n-1)}\sum_{i=1}^{n}\sum_{j\neq
i}^{n}\h e_{i(1)}K_h(\h B(\h q)^{\top}w_i-\h B(\h q)^{\top}w_j)\h e_{j(2)}
\eenr
to perform the test. We shall prove that the asymptotic bias of $V_n$ vanishes, but its
asymptotic variance gets larger than that of $\ti V_n$.
Note that $\ti V_n$ and $V_n$ are non-standardized, the standardizing constants will be
specified in Section~3. Here, we mention a significant feature of both of these statistics,
which is that their asymptotic behavior is like that of a test statistic with one-dimensional
covariate $X$, i.e., their consistency rate is $1/\sqrt {nh^{1/2}}$, which in turn  greatly
alleviates the dimensionality issue.

From the above construction, it is obvious that estimating adaptively the matrix $B$ under the
null and alternative hypothesis plays a crucial role for dimension reduction.
The next subsection is devoted to this issue.

\subsection{Estimation of  B and $\b$}
To achieve the adaptation property of the estimators of $B$ and $\b$  mentioned above,
the key is to derive an estimator of $B$ up to an $q\times q$ orthonormal matrix $C$ without depending on the assumed models  under the null and alternative hypotheses.
With measurement errors, Carroll and Li (1992) extended sliced  inverse regression (SIR, Li 1991) to errors-in-variables regression models.
Lue (2004) extended the principal Hessian directions (pHd, Li 1992) method to the surrogate problem.
Li and Yin (2007) established a general invariance law between the surrogate and the original dimension reduction spaces when $X$ and $U$ are jointly multivariate normal.
If $X$ or $U$ is not normally distributed, they suggested an approximation based on the results of Hall and Li (1993).
See also Zhang, Zhu and Zhu (2014).

As the discretization-expectation estimation method (DEE) of Zhu et al.\,\,(2010a) is simple to implement without selecting the number of slices, we adopt it to errors-in-variables models when SIR is used.
Write $S_{Y|X}$ as the central subspace that is the intersection of all column spaces spanned by the columns of $B$ that makes $Y$ conditionally independent of $X$, given $B^{\top}X$, i.e., $Y\hDash X|B^{\top}X$.
This means that identifying $S_{Y|X}$ is equivalent to identifying a base matrix $\ti B$ that is equal to $BC^{\top}$ for a $q\times q$ orthogonal matrix $C$.
Note that the function $G$ is unknown in the alternative. We can rewrite $G(B^{\top}X)$ as $\ti G(\ti B^{\top}X)$.
In other words, identifying $\ti B$ is enough for model identification.
Without notational confusion, we write $\ti B=B$ throughout the rest of this paper.

To extend the DEE method to the setting with measurement errors, we first give a very brief review.
Assume that $\mb{Cov}(X)$ is the identity matrix.
As is known, SIR is fully dependent on the reverse  regression function $E(X|Y)$ such that we can consider the eigen-decomposition of its covariance matrix $\Cov(E(X|Y))$.
The eigen vectors associated with nonzero eigen values of this matrix form the base matrix $B$.
SIR-based DEE uses the matrix $\L=E\{\Cov(E(X|\ti Y(T)))\}$  as the target matrix, where $\ti Y(t)=I(Y\le t)$, $t \in \R$ and $T$ is an independent copy of  $Y$.
Because the measurement error $U$ is independent of $Y$, and thus, when $X$ is replaced by $W$, at the population level, nothing is changed about eigen-decomposition and eigen vectors.
We use surrogate predictors $\Cov(X,W)\Si_{W}^{-1}W$, which forms the least squares prediction of X when W is given.
Carroll and Li (1992) pointed out that sliced inverse regression (SIR) with the surrogate predictors can produce consistent estimators of $S_{Y|X}$.
In other words, all steps of estimation are exactly the same as those in the without measurement errors set up.
The reader can refer to  Zhu et al.\,\,(2010a) for more details.

When we use data to construct an estimate $\L_n$ of $\L$, we can then obtain an estimate $\hat{B}(\hat{q})$ of $B$, which consists of the $\h q$ eigenvectors of $\L_n$ with non-zero eigenvalues, where $\h q$ is defined as follows,  using the BIC type criterion proposed by Zhu et al.\,\,(2006).
Let $\h \la_1\geq\h \la_2\geq\cdots\geq\h \la_p$ be the eigen values of the matrix  $\L_n$ in descending order.
An estimate $\h q$ of $q$ is given by
\benr\label{qbic}
\h q=\arg\max_{l=1,\cdots,p}\left\{\frac{n}{2}\times\frac{\sum_{i=1}^{l}
\{\log(\h \la_i+1)-\h \la_i\}}{\sum_{i=1}^{p}
\{\log(\h \la_i+1)-\h \la_i\}}-2\times D_n
\times \frac{l(l+1)}{2p}\right\},
\eenr
where $D_n$ is a sequence of constants not depending on the data.
Here we take $D_n=n^{1/2}$.

The following consistency results can be obtained from Zhu et al.\,\,(2010a).
\begin{prop}\label{pro1}
Suppose the assumptions in Zhu et al.\,\,(2010a)
hold and $N/n\to \la$. Then the following hold. \\
(1). Under $H_0$, $P(\h q=1)\to 1$, and $B$ is a vector proportional to $\b$. Moreover,
\benr\lel{bq}
 \h B(\h q)-B&=&O_p(1/\sqrt n), \quad 0<\la\le \iny,\\
 &=&O_p(1/\sqrt N), \hs .8cm \la=0. \nn
\eenr
(2). Under $H_1$, $P(\h q=q)\to 1$, $B$ is a $p\times q$ orthonormal matrix
and $\h B(\h q)$ satisfies (\r{bq}).
\end{prop}

There are various estimators of $\b$ for EIVs models available in the literature.
Here we shall focus on the estimators proposed by Lee and Sepanski (1995) for
linear and nonlinear EIVs regression models.
Their estimator under the null hypothesis is
\benrr
\h \b=\arg\min_{\b} ({\mathbf Y}-{\mathbf D}({\mathbf D}_v^{\top}
{\mathbf D}_v)^{-1}
{\mathbf D}_v g(\bX_v\b))^{\top}(Y-{\mathbf D}({\mathbf
D}_v^{\top}{\mathbf D}_v)^{-1}
{\mathbf D}_v g(\bX_v\b))
\eenrr
where ${\mathbf X}_v$ is the $N\times p$ matrix whose sth row is $ \ti
x_s^T, s=1,\cdots, N$, ${\mathbf Y} $ is a $n\times 1$ vector, and
$g(\bX_v\b)$ represents  $N\times 1$ vector $[g(\b^{\top} \ti
x_{1}),\cdots,g(\b^{\top} \ti x_N)]^{\top}$.
The matrices ${\mathbf D}$ and ${\mathbf D}_v$ are design matrices according to
$g(\cdot)$. More precisely, ${\mathbf D}$ is the $n\times k$ matrix whose
$i$-th row denoted by $\bar{w}^{'}_i$, is a vector consisting of polynomials of
$w_i$, while ${\mathbf D}_v$ is the corresponding matrix of validation data,
whose $s$-th row $\bar{w}_s$ is a vector consisting of polynomials of $\ti w_s$.
For linear model, $\bar{w}_i=w_i$ and $\bar{w}_s=\ti w_s$.
For nonlinear model, we let $\bar{w}_i$($\bar w_s$) be the
vector consisting of a constant and the first two order polynomials of
$w_i$($\ti w_s$). Lee and Sepanski (1995) assume that $\lim \sqrt{n/N}$ exists.
They show that if this limit is non-negative and finite then $\h \b$ is root-$n$
consistent for $\b$, and  if $\lim \sqrt{n/N}=\iny$, then $\h \b$ is a root-$N$
consistent for $\b$. More precisely,
we have the following proposition.
\begin{prop}\label{pro2}
Suppose the assumptions for Proposition 2.2 in Lee and Sepanski (1995) hold.\\
{\rm (1)}. Suppose in addition $H_0$ holds and $N/n \to \la$. Then for $0< \la\le \iny$,
 $\sqrt{n}(\h \b-\b)=O_p(1)$, while for $\la=0$, $\sqrt{N}(\h \b-\b)=O_p(1)$.\\
{\rm (2)}. In addition, suppose the following sequence of  local alternatives holds, where $C_n\to 0$.
\benrr
H_{1n}: \mu(x)=g(\b^{\top}x)+C_{n}G(x).
\eenrr
Then
\benrr
\h \b-\b_0 &=& C_n \left\{E[g'(\b^{\top}X)X\bar{W}^{\top}]E^{-1}[\bar{W}\bar{W}^{\top}]E[g'(\b^{\top}X)\bar{W}X^{\top}]\right\}^{-1}\\
                    && \hs .5cm \times E[g'(\b^{\top}X)X\bar{W}^{\top}]E^{-1}[\bar{W}\bar{W}^{\top}]
                    E[\bar{W}G(X)](1+o_p(1))\\
                    && +O_{p}(1/\sqrt{n})+O_{p}(1/\sqrt{N}).
\eenrr
where $\bar W$ is a vector consist of polynomials of $W$ and $g'(t)$ is the derivative of $g(t)$ with respect to $t$.
\end{prop}

\section{Asymptotic distributions}
\subsection{Limiting null distribution}
In this section, we will establish the asymptotic null distribution of the
proposed test statistics  $\ti V_n$ in (\ref{7}) and
$V_{n}$ in (\ref{8}). Define

\benr\lel{defs1}
&&Z=B^{\top}W, \quad \si^2(Z)=E[e^2|Z], \quad \D(Z)=E[G(B^{\top}X)|Z],\\
&&\eta=g(\b^{\top}X)-r(\b^{\top}W,\b), \quad \xi^2(Z)=E[\eta^2|Z], \nn
\eenr
where $e$ is defined in (\ref{res}).
Write $Z$ as $\ti Z$, when $W$ is replaced by validation data $\ti W$.

To proceed further we now state the assumptions needed here.

\noi{\bf Assumptions:}\\[.1cm]
\noi (f). The support $\mathcal{C}$ of $Z$ is a compact subset of the support of
$\ti Z$ and bounded away from the boundary of the support of $\ti Z$. The density
$f$ of $Z$ has bounded partial derivatives up to order $\ell\ge 1$ and satisfies
\benrr
0<\inf_{z\in \mathcal{C}}f(z)\leq\sup_{z\in \mathcal{C}}f(z)<\iny.
\eenrr
\noi (g). $g(\b^{\top}x)$ is a measurable function of $x$ for each $\b$ and is differentiable in $\b$ up to order $\ell+1$, and $E\big\|\frac{\partial g(\b_0^{\top}X)}{\partial \b}\big \|^2<\iny$.
\vs .15cm
\noi (r).  The function $r(\b^{\top} w, \b)$ has bounded partial derivatives with respect to $\b^T w$  up to order $\ell+1$, and  $E[r^2(\b^{\top} W, \b)]<\iny$, $\b \in \R^p$.
\vs .15cm
\noi (G). $E[\D^2(Z)]<\iny$, $E[(G(B^{\top}X)-\D(Z))^4]<\iny$, and $\D(z)$ has bounded partial derivatives up to order $\ell$.
\vs .15cm
\noi (W). $\max_{1\le k\le p} E[W^2_{(k)}|Z]<\iny$, $W_{(k)}$ represents the $k$-th coordinate of $W$, $k=1,\cdots,p$.
\vs .15cm
\noi (e). $E[(\si^2(Z))^2]<\iny$, $E[(\xi^2(Z))^2]<\iny$, and $\si^2(z)$ and $\xi^2(z)$ are uniformly continuous functions.
\vs .15cm
\noi (K). $K$ is a spherically symmetric and continuous kernel function with bounded support and  of order $\ell$, having all derivatives bounded.
\vs .15cm
\noi (M). $M$ is a symmetric and continuous kernel function with bounded support and  of order $\ell$, having all derivatives bounded.
\vs .15cm
\noi (h1). $h\to  0$, $v_N \to  0$, $v_N/h\to 0$.
\vs .15cm
\noi (h2). $h\to  0$, $v_N \to  0$, $h^{4}/v_N^{5}\to 0$.
\vs .15cm
\noi (h3). $nh^2\to \iny$, $Nv_N^2\to \iny$, $n v_N^{2\ell}\to 0$ and $nhv_N/N\to 0$.
\vs .15cm
\noi (h4). $nh^{5/2}\to  \iny$, $Nv_N^2\to \iny$, $nv_N^{2\ell}\to 0$ and $nhv_N/N\to 0$.
\vs .15cm
\noi (h5). $nh\to \iny$, $Nh^2\to \iny$, $Nv_N^{1/2}/(nh^{1/2})\to 0$ and $Nv_N^{1/2+2\ell}\to 0$.
\vs .15cm
\noi (h6). $nh^{q}\to \iny$, $Nv_N\to \iny$.
\hs .2cm

The positive integer $\ell$ in all of the above assumptions is the same as in the assumption (f).
For the consistency of $\h \b$ and $\h B (\h q)$, some additional conditions
are also needed. The reader can refer to Lee and Sepanski (1995) and Zhu et al.\,\,(2010a) for
more details.

\begin{rema}
{\rm Conditions (g), (r), (W), (e) are very common for
the asymptotic normality of the proposed test  statistics.
The lower bound assumption on $f$ is typically designed for the
nonparametric estimation of the corresponding regression function $r(\b^{\top}W, \b)$
and the conditional mean $E[e|Z]$.
This is a commonly used condition.
In assumption (h6), $nh^{q}\to \iny$ is to ensure the consistency in quadratic
mean of kernel density estimator under some global alternative.
If $v_N/h\to  0$, some convolution of kernel functions can be approximated by
kernel function.  If $N/n\to  \iny$ or a finite constant, this condition is
easily satisfied. We choose $v_N=O((N/2)^{-2/5})$ in the simulation studies later.
But when $N/n\to  0$, the condition is changed to $h/v_N\to  0$.}
\end{rema}

To proceed further, we need some more notation as follows:
\benr\lel{defs2}
z_i=B^{\top}w_i, \quad g_i=g(\b^{\top}x_i), \quad r_i=r(\b^{\top}w_i, \b),\quad
\eta_i=g_i-r_i.
\eenr
Write $\ti z_s$, $\ti g_s$, $\ti r_s$ and $\ti \eta_s$ for the entities in
(\r{defs2}) when $w_i$ is replaced by validation data $\ti w_s$ in there. When $\b$
and $B$ are respectively replaced by their estimators
$\h \b$ and $\h B(\h q)$ in the above definitions, write the respective $\h z_i$, $\h
g_i$, $\h r_i$ and  $\h \eta_i$ for $z_i$, $g_i$, $r_i$ and $\eta_i$, and similarly
write the respective $\h {\ti z}_s$, $\h {\ti g}_i$, $\h {\ti r}_i$ and
$\h {\ti \eta}_i$ for $\ti z_i$, $\ti g_i$, $\ti r_i$ and $\ti \eta_i$.

To state the next theorem we need to define
\benr\lel{si123}
&& \mu=K(0)E[\xi^{2}(z)]/(Nh), \qquad
\t_{1}=2\int K^{2}(u)du\int(\si^{2}(z))^{2}f^{2}(z)dz, \\
&&  \t_{2}=\int K^{2}(u)du\int\si^{2}(z)\xi^{2}(z)f^{2}(z)dz,\quad
\t_{3}=2\int K^{2}(u)du\int(\xi^{2}(z))^2f^{2}(z)dz. \nn
\eenr
where $\si^{2}(\cdot)$ and $\xi^{2}(\cdot)$ are defined in (\ref{defs1}) and $f$ is
the density of $Z=B^{\top}W$.
Consistent estimates of $\Si_i, i=1,2,3$ under $H_0$ are given by
\benr\lel{hsi123}
&& \h \t_1= \frac{2}{n(n-1)}\sum_{i=1}^{n}\sum_{j\neq i}^{n}\frac{1}{h^{\h q}}K^{2}
(\frac{\h z_i-\h z_j}{h})\h e_i^2\h e_j^2, \quad
\h \t_2=\frac{1}{nN}\sum_{i=1}^{n}\sum_{s=1}^{N}\frac{1}{h^{\h q}}K^{2}
(\frac{\h z_i-\hat{\ti z}_s}{h}) \h e_{i}^{2}\h{\ti \eta}_s^2\\
&& \h \t_3= \frac{2}{N(N-1)}\sum_{s=1}^{N}\sum_{s^{'}\neq s}^{N}
         \frac{1}{h^{\h q}}K^{2}(\frac{\h{\ti z}_s-\h{\ti z}_{s^{'}}}{h})\h{\ti \eta}_{s}^2\h{\ti \eta}_{s{'}}^2. \nn
\eenr
 We are now ready to state
\begin{theo}\label{theo3.1}
Suppose $H_{0}$ and the conditions (f), (g), (r), (W), (e), (K), (M), (h1) and
(h3) hold, and that
$N/n\to\la$, $0<\la\le \iny$.
Then
$
nh^{1/2}\big(\ti V_n-\mu\big)\to_D N(0,\ti \t),
$
where
\benrr
\ti \t&=&\t_{1}+\frac{2}{\la}\t_2+\frac{1}{\la^2}\t_3, \quad
0<\la<\iny,\\
&=& \t_1, \hs 1.40in \la=\iny.
\eenrr
\end{theo}
Here, consistent estimators of  $\mu$ and $\t$ under $H_0$ are given by
\benrr
\h \mu =\frac{1}{N^2h}K(0)\sum_{s=1}^{N}\h{\ti \eta}_s^{2}, \qquad
\hat{\ti \t}=
\h \t_1+\frac{2}{\la}
\h \t_2+\frac{1}{\la^2}\h \t_3, \quad 0<\la<\iny,
\eenrr
with $\h \t_i$'s as in (\r{hsi123}). The $\ti V_n$ test rejects $H_0$ whenever
$\ti V_n > \hat{\ti \t}^{1/2}(nh^{1/2})^{-1} z_\al + \h \m$,
where $z_{\al}$ is the upper $100(1-\al)\%$ quantile of the standard normal distribution.

The above theorem shows that the asymptotic variance of $\ti V_n$ consists
of the three parts when $0<\la<\iny$.
The part $\t_1$ reflects the variation in the regression
model, $\t_3$ is the variation caused by the measurement
error while the part $\t_2$ is the intersection of the variation due to the regression
model and measurement
error.

The next result gives the asymptotic null distribution of the $V_n$ statistic of (\ref{8}).
As  can be seen from this result, $V_n$ does not have any
asymptotic bias.
\begin{theo}\label{theo3.2} Under the conditions of Theorem \r{theo3.1},
$
nh^{1/2}V_{n}\to_D N(0,\t),
$
where
\benrr
\t &=&\t_{1}+\frac{4}{\la}\t_{2}+\frac{2}{\la^2}\t_{3},
\quad 0<\la<\iny,\\
&=&\t_1, \hs 1.40 in \la=\iny,
\eenrr
where $\t_i$, $i=1,2,3,$ are as in (\r{si123}).
\end{theo}

To studentize $V_n$, we use the following consistent estimate of $\t$
in the case $0<\la<\iny$.
\benrr
\h \t
 =&\frac{2}{n(n-1)}\sum_{i=1}^{n}\sum_{j\neq i}^{n}\frac{1}{h^{\h q}}
          K^{2}(\frac{\h z_i-\h z_j}{h})\h e_{i(1)}^{2}\h e_{j(2)}^{2}
+\frac{4}{\la nN}\sum_{i=1}^{n}\sum_{s=N/2+1}^{N}\frac{1}{h^{\h q}}
K^{2}(\frac{\h z_i-\h{\ti z}_s}{h})\h e_{i(1)}^{2}\h{\ti \eta}_s^2 \\
& +\frac{4}{\la nN}\sum_{i=1}^{n}\sum_{t=1}^{N/2}\frac{1}{h^{\h q}}
K^{2}(\frac{\h z_i-\h{\ti z}_t}{h})\h e_{i(2)}^{2}\h{\ti \eta}_{t}^2
 +\frac{16}{\la^2N^{2}}\sum_{t=1}^{N/2}
           \sum_{s=N/2+1}^{N}\frac{1}{h^{\h q}}K^{2}(\frac{\h{\ti z}_s-\h{\ti z}_t}{h})
           \h{\ti \eta}_s^2\h{\ti \eta}_{t}^2,
\eenrr
where $s$ and $t$ are indices of the two sets of validation data respectively,
$\h \eta_t$ or $\h \eta_s$ is estimated by the other half of validation data.
That is, $\h{\ti \eta}_t=g(\h \b^{\top}\ti x_t)-\h r_{(2)}(\h \b^{\top}\ti w_t, \h \b)$, $t=1,\cdots, N/2$ and $\h{\ti \eta}_s=g(\h \b^{\top}\ti x_s)-\h r_{(1)}(\h \b^{\top}\ti w_s, \h \b)$, $s=N/2+1,\cdots, N,$ where $\h r_{(1)}$ and $\h r_{(2)}$ are defined in (\ref{r12hat}).
The standardized test statistic is
\benrr
T_{n}&=&\h \t^{-1/2}nh^{1/2}V_n,
\quad 0<\la <\iny,\\
&=& \hat \t_1^{-1/2}nh^{1/2}V_n, \hs .8cm \la=\iny,
\eenrr
where $\h \t_1$ is as in (\r{hsi123}).
According to the Slusky theorem, $T_n$ is asymptotically standard normal.
At the significance level  $\al$, the null hypothesis is rejected when $T_n>z_\al$.
For large $\la$, the terms about $\t_2$ and $\t_3$  vanish in the asymptotic variance, and thus, the estimated variance $\h \t$ is replaced by $\h \t_1$.
\begin{rema}
{\rm A significant feature of this test is that we only need to use the standardizing sequence  $nh^{1/2}$, which is the same as the one used in the classical local smoothing tests when  $X$ is one-dimensional.
This shows that the test statistic has a much faster convergence rate to its limit compared to some of the classical tests that have the rate of order $nh^{p/2}$.
This greatly assists in maintaining the significance level of this test in finite samples when its asymptotic null distribution is used to determine the critical values for its implementation.}
\end{rema}

When $N/n\to \la=0$, the standardizing constant will be different because of the plug-in estimate $\h r(\cdot)$ of the function $r(\cdot)$, as is evidenced by the following theorem.
\begin{theo}\label{theo3.3}
Suppose $H_{0}$ and the above conditions (f), (g), (r), (W), (e), (K), (M), (h2), (h5) hold and that $N/n\to 0$. Then
$
Nv_N^{1/2}\{\ti V_n-\nu\}\to_D N(0,\ti \t),$ \, $ Nv_N^{1/2}V_{n}\to_D N(0,\t),
$
where $\nu = \|\b\|(v_N N)^{-1}\int M^2(u) du \, E[\xi^2(Z)],$
\, $\t:= 2\ti \t,$ and
\benrr
&&\ti \t=2 \|\b\|\int \Big(\int M(u)M(u+v)du \Big)^2 dv \int (\xi^2(z))^2 f^2(z) dz.
\eenrr
\end{theo}

\subsection{Asymptotic Power}
In this section, we assume $N/n\to \la$, $\la$ a positive constant and investigate the asymptotic
properties of the  test statistic $V_n$ under global  and local alternatives. This is because
the asymptotic properties can be much more easily derived than those for $\ti V_n$.
Consider a sequence of  alternatives
\benr\label{localalterB}
H_{1n}:\m(x)=g(\b^{\top}x)+C_{n}G(B^{\top}x), \quad x\in \R^p,
\eenr
where
$G(\cdot)$ satisfies $E(G^{2}(B^{\top}X))<\iny$ and $\b$ is a column of $B$.
When $C_n$ is a fixed constant, the alternative is a global alternative and when $C_n=n^{-1/2}h^{-1/4}$ tends to zero, $H_{1n}$ specify the local alternatives of interest here.
Note that the asymptotic properties of the estimates $\h B(\h q)$ and $\h \b$ will affect the behavior of the test statistic $V_n$.
The asymptotic results of $\h \b$ have been illustrated in Proposition~ \ref{pro2}.
Thus, we  discuss the  result  about the consistency of $\h q$ here.
Under the local alternatives, it is no longer consistent for the dimension $q$.
\begin{theo}\label{th3.4}
Suppose the conditions in Zhu et. al (2010a) hold. Under $H_{1n}$ of (\ref{localalterB}) with
 $C_n=n^{-1/2}h^{-1/4}\to 0$, $P(\h q= 1)\to 1$.
\end{theo}

However, this inconsistency does not hurt the power performance of the test.
We will see below in a finite sample simulation study  that the test can be much more powerful than the classical local smoothing tests in the literature.
\begin{theo}\label{th3.5}
Under the alternatives of (\ref{localalterB}), the following results are hold:\\
(i)Suppose (f), (g), (r), (G), (W), (e), (K), (M), (h1) and (h6) hold. Under the global alternative with fixed $C_n$,
\benr\lel{cons}
V_{n}/\h \t\to V>0.
\eenr
(ii) Suppose (f), (g), (r), (G), (W), (e), (K), (M), (h1) and (h4) hold. Then, under the local alternatives
$H_{1n}$ with  $C_{n}=n^{-1/2}h^{-1/4}$,
$nh^{1/2} V_{n} \to_D N(\D,\t)$,
where $\t$ is given in Theorem~\ref{theo3.2} and
$
\D=E\left[\{\D(Z)-E[g{'}(\b_0^{\top}X)X^{\top}|Z]H(\b_0)\}^2f(Z)\right].
$
\end{theo}
\begin{rema}
{\rm The result (\r{cons}) implies the consistency of the $T_n$ test gainst the class of the above fixed alternative.
It also implies that under the global alternatives, the test statistic can diverge to infinity at a much faster rate than the existing local smoothing tests in the literature can achieve such as Zheng's test (1996), which has the consistency rate of the order $1/(nh^{p/2})$.
The test can also detect the local alternatives distinct from the null at the rate of order $1/\sqrt {nh^{1/2}}$ while the classical ones can only detect those alternatives converging to the null at the rate of order $1/\sqrt {nh^{p/2}}$.}
\end{rema}

\section{Numerical studies}
This section presents four simulation studies to examine the performance of the
proposed test ($T_{n}$). To compare with existing tests, we consider Zheng's (1996)
test ($T_{n}^{Zh}$) adapted to the errors-in-variables settings and Song's (2009) test
($T^{S}_{n}$) as the competitors. The adapted Zheng's test is the same as our test
except that $B^{\top}W$ is replaced by the original $W$. This is a typical local
smoothing test. Song's test is a score type test and is designed for EIVs models with
validation data. Consider the linear regression models under the null hypothesis.
In the simulation study~1 below, the matrix $B$ is equal to $\b$ and thus, the model is
a parametric single index. The dimension of $X$ is respectively $p=2$ and $8$.
Note that our test fully uses the information under the null hypothesis that only
relates to a single index $\b$. In addition, we run simulation studies of the test
$\ti T_n$ based on the statistic $\ti V_n$ of Theorem~\ref{theo3.1}
when $0<\la <\iny$, and illustrate its weakness.
The purpose of Study~2 is to confirm that the proposed test $T_n$ is not a directional
test by assuming $B=(\b_1, \b_2)$ with $q=2$ under the alternative hypothesis.
Study 3 is designed to examine the finite sample performance when $N<n$ and $N>n$.
Study 4 considers four nonlinear models. All simulations are based on 2000 replications.

Recall that the tests $T_{n}$ and $T_{n}^{Zh}$ are based on the estimates of the quantities that are zero under the null and positive under the alternative.
Because of the asymptotic normality, the rejection regions of $\ti V_n$, $T_{n}$ and $T_{n}^{Zh}$ are one-sided: $\{\ti V_n >
\hat{\ti \t}^{1/2}(nh^{1/2})^{-1} 1.65 + \h \m \}$, $\{T_{n}>1.65\}$ and $\{T_{n}^{Zh}>1.65\}$ at the $0.05$ level of significance.
The reported size and power are computed by $\#\{T_{n}>1.65\}/2000$.
For $T^{S}_{n}$, the rejection region is two sided and the reported size and power are computed by $\#\{|T^S_{n}|>1.96\}/2000$.
Throughout the simulation studies, $X$ is taken to be multivariate normal with mean zero and covariance matrices
$\Si_{1}=I_{p\times p}$ and $\Si_{2}=(0.3^{|i-j|})_{p\times p}$.
The regression model error $\vep$ follows standard normal distribution, while the  measurement error $U\sim N(0, 0.5)$.
The kernel function is $K(u)=\frac{15}{16}(1-u^{2})^{2}I(|u|\leq1)$ which is a second-order symmetric kernel and $M(u)=K(u)$.

\noindent {\bf  Bandwidth selection.} \, As the tests involve bandwidth selection in the kernel estimation, we run a simulation to empirically select the bandwidths for the three tests in the comparison.
Because the significance level maintainance is important, we then select bandwidths such that the tests can have empirical sizes close to the significance level and retain the use under other models.
To this end, we use a simple model to select them and to check whether they can be used
in general. In our test, there are two bandwidths.
As is well known, the optimal bandwidth in hypothesis testing is still an outstanding
problem, but the optimal rate of the bandwidth in kernel estimation is  $n^{-1/(4+q)}$
where $n$ is the sample size.
We then adopt its rate with a search for the constant $c_1$ in
$h=c_{1}n^{-1/(4+\h q)}$. Similarly, for the kernel estimator of the function
$r(\b^{\top}W,\b)$, we choose the window  width
$v_N=c_{2}(N/2)^{-2/5}$, because we halved the validation data set of size $N$.
For $\ti T_n$, $v_N$ is $c_2 N^{-2/5}$.
To select proper bandwidths, we tried different bandwidths to investigate their impact
on the empirical size. To reduce the computational burden, we consider $c_1=c_2=c$ to
see whether such selections can offer bandwidths for general use.
The selection is based on hypothetical models as the primary target is to maintain the
significance level. Thus, we compute the empirical size at every equal gird point  $c=(i-1)/10$ for $i=1, \cdots 21$.
In Figure~1, we report the empirical sizes associated with different bandwidths when the regression model is
$ \m(x)=\b^{\top}x$ and $p=2,8$, $n=100, 200$, $N=4\times n$, and the covariance matrix of $X$ is
$\Si_{1}$.
We can see that the test is not very sensitive to the bandwidth and a value of $c=1.6$ may be a good choice
for both $T_n$ and $\ti T_n$.
For the adapted Zheng's test, there are also two bandwidths to be selected.
As the optimal rate for the kernel estimation is $h=c_1n^{-1/(4+p)}$, we then also consider $c_1=c_2=c$.
We found that to maintain the significance level, the bandwidths must be with larger $c$.
The initial selection provides us an idea to choose a good bandwidth within the equal grid points as $c=2.5+(i-1)/10$ for $i=1, \cdots 21$.
The results are also reported in Figure~1.
As for Song's score test, only one bandwidth is required.
We also found a larger bandwidth is required.
Set the bandwidth as $v_N=cN^{-1/(4+p)}$ and search for the proper $c$ within the equal grid points as $c=1+(i-1)/10$ for $i=1, \cdots 21$.
The reported curves are in Figure~1.
\begin{center}
Figure~ 1. about here
\end{center}
We can see that the empirical sizes of $T_n$ are not sensitively affected by the bandwidths selected.
The curves of empirical size under $p=2$ and $p=8$ are almost coincident.
While the empirical size of $\ti T_n$ is slightly effected by dimensionality, but it
is still more robust than that of $T_n^{Zh}$ and $T_n^S$.
A value of $c=1.6$ is worthy of recommendation for both, $T_n$ and $\ti T_n$.
However, the empirical sizes of $T_{n}^{Zh}$ and $T_{n}^{S}$ associated with the
bandwidths are not as robust as that of $T_{n}$.
The empirical sizes show the efficient bandwidth changes as $p$ increase.
When $p$ is small, a small $h$ can keep the theoretical size.
As $p$ increase, a larger $h$ is necessary.
This phenomenon is particularly serious for $T_n^{Zh}$.
For the bandwidths of $T_n^{Zh}$, $c=3.9$ is appropriate.
Finally,  $c=2.2$ seems to be proper for $T_n^{S}$.

{\bf Study 1.} The data are generated from the following model:
\benrr
&&H_{11}:\, \, \m(x)=\b^{\top}x+a\, (\b^{\top}x)^2, \\
&& H_{12}:\, \,\m(x)=\b^{\top}x+a\,\exp(-(\b^{\top}x)^{2}/2), \\
&& H_{13}:\, \,\m(x)= \b^{\top}x+2a\cos(0.6\pi\b^{\top}x).
\eenrr
The case of $a= 0$ corresponds to the null hypothesis and $a\not = 0$ to the alternatives.
In other words, both the hypothetical and  alternative models have a single index $B=c\b$.
Models under $H_{11}$ and $H_{12}$ represent low frequency alternatives while $H_{13}$ is an example of high frequency alternative.
In $H_{11}$ and $H_{12}$, the alternative parts $(\b^{\top}x)^2$ and $\exp(-(\b^{\top}x)^{2}/2$ always exist for any nonzero $a$. While for $H_{13}$, the alternative part $\cos(0.6\pi\b^{\top}x)$ appears and disappears periodically for $a\neq 0$, which makes the bandwidth selection process even more challenging. Because a large bandwidth selected to maintain significance level may make the test obtuse to high frequency alternatives.
The dimension $p$ equals $2$ and $8$ such that we can check the impact from the dimensionality.
Let $\b=(1,1,\cdots,1)^{\top}/\sqrt{p}$.
The number of validation data is $N=4n$.
The simulation results are presented in Tables~1, 2 and 3.
\begin{center}
Tables~1-3 about here
\end{center}

From these tables we see  that when $p=2$, $T^{S}_{n}$ performs very well.
This is expected when the dimension is low or moderate, because the consistency rate of this test is $1/\sqrt n$.
Also, when $p$ is small, $T_{n}^{Zh}$ is comparable to $T_n$ as both are local smoothing tests.
When the dimension increases, $T_{n}^{Zh}$ and $T^S_n$ are however severely impacted by the dimensionality. The test
$T_{n}^{Zh}$ behaves much worse. Especially, when $p=8$, it breaks down for $n=100$ and regains its power as $n$ increase.
The test $T^S_n$ is also affected by the dimensionality because the residuals contain nonparametric estimation by local smoothing technique.
Its powers decrease both for small and large sample size.
On the other hand, the dimension-reduction adaptive-to-model test $T_n$ does not suffer from the curse of dimensionality in the limited simulation studies presented here.
When $p$ is large, $T_n$  performs better than $T^S_n$.
The finite sample power of the $T^S_n$ test is poor against the alternatives $H_{13}$ for both the cases $p=2$ and $p=8$.
This may be due to the fact that $T^S_n$ is a directional test.
We illustrate this problem in the next study.

The comparison between $T_n$ and $\ti T_n$ is another purpose of this study.
We find that the empirical power of $\ti T_n$ is slightly higher than
that of $T_n$, but the size of $\ti T_n$ also tends to be slightly larger, even
when $n=200$ and $p=2$. Although $\ti T_n$ has bias, but each residual in $\ti T_n$ is
estimated by all validation data which is more precise with smaller variance than that
of $T_n$ derived by half validation dat.
We can then conclude, based on this limited simulation, the test $\ti T_n$ is slightly
more liberal than the bias-corrected test $T_n$, but
also slightly more powerful.
These two tests are competitive. Therefore, in the following simulation studies, we only report the results about $T_n$ to save space.

{\bf Study 2.}
In this study, we aim to design a simulation study to check that the dimension-reduction model-adaptive test $T_n$ is not a directional test, while Song's test $T^{S}_{n}$ is.
The data are generated from the following model:
\benrr
&&H_{14}:\, \, \m(x)=\b_{1}^{\top}x+a(\b_{2}^{\top}x)^2,\qquad
H_{15}:\, \, \m(x)=2\b_{1}^{\top}x+a(2\b_{2}^{\top}x)^3.
\eenrr
Here also, $a=0$ corresponds to the null hypothesis and $a\ne 0$ to the alternatives.
The matrix $B=(\b_1, \b_2)$ and then the structural dimension $q$ under the alternative
is $2$. Let $p=4$, $\b_1=(1,1,0,0)^{\top}/2$ and $\b_2=(0,0,1,1)^{\top}/2$.
The number of validation data is $N=4\times n$.
The simulation results are presented in Table~4.
From these results, we first observe that $T^S_n$ has good
performance under $H_{14}$, which coincides with that in Study~1.
However, the poor performance under $H_{15}$ shows that $T^{S}_{n}$ is a directional
test as this alternative cannot be detected by it at all.
At population level, we can see that the conditional expectation of the residual is
equal to zero under this alternative. In this case, $T_n$ still works well.
This lends support to the claim that $T_n$ is an omnibus test.
\begin{center}
Tables~4 about here
\end{center}

{\bf Study 3.}
In this study, we aim to explore the impact of the estimation of $r(\cdot)$ on the
performance of the proposed tests.
Small $\la=\lim(N/n)$ means that there are not many validation data available and large
$\la$ means the estimator  $\h r(\cdot)$ is very close to the true function $r(\cdot)$.
For this purpose, consider $N/n=0.1, 0.5, 4, 8$. We only choose these ratios because if
$\la$ is either too small or too large, we need to have too large sample size or too
large size of validation data.
These are practically not possible.
From Theorem~\ref{theo3.3}, we know that when $\la$ is small, we can have a test with
simpler limiting variance. Write the related test as $T_n^{(1)}$.
From Theorem~\ref{theo3.2}, $\la = \iny$ case, we can also have a test
for  large $N/n$. Write it as $T_n^{(2)}$.
To examine whether these two variants of the test $T_n$ work or not, we generate data from the model $H_{11}$ in {\bf Study 1}.
When the size of validation data is such that $N/n=0.1, 0.5$, $T_n^{(1)}$ is used, and when  $N/n=4,8$, $T_n^{(2)}$ is  applied.
As $T_n^{(1)}$ is a test with very different convergence rate, we then also need to choose bandwidths suitable for it.
Similarly as the above, we also search for the bandwidths at the rates $v_N=c_1(N/2)^{-1/3}$ and $h=c_2n^{-1/(2+\h q)}$.
Let $c_1=c_2=c$.
We found that $c=2$ is a good choice.
For $T_n^{(2)}$, only the asymptotic variance changes, we then still use the same bandwidths as before.
When $\la=0.1,0.5$, we then use larger sample size of validation data $N=100, 200$, otherwise, $N$ is too small to make the tests well performed.
The simulation results are presented in Table~5.
\begin{center}
Table~5 about here
\end{center}

From Table~5, we have the following two observations.
First, for $\la=0.1$, $T_n$ is more conservative with lower  power
than $T_n^{(1)}$. This seems to say, $T_n$ is less sensitive to the alternative
model than $T_n^{(1)}$. This phenomenon would come from the improper selection of
bandwidths for $T_n$ because Conditions (h1) and (h2) assure that the
consistency of $T_n$ and $T_{n}^{(1)}$ require different ratios of $h$
and $v_N$. Thus, when $N/n$ is very small, $T_n^{(1)}$ seems to be
a better choice than $T_n$. But when $\la$ is closed to 1, $T_n^{(1)}$ cannot
maintain the significance level well. Secondly, $T_{n}^{(2)}$  has very
slightly higher empirical size and power than $T_n$. Overall, the performances
of $T_{n}^{(2)}$  is very similar to that of $T_n$.
Therefore, when the size of validation data $N$ is reasonably large, and the ratio
$N/n$ is large, $T_{n}^{(2)}$ would be applicable.
Also, from the simulations we see that although $T_n^{(1)}$ can be used,
it does not maintain the finite sample significance level as well as the $T_{n}$
test does. Thus, when the ratio $N/n$ is not too small, we recommend the test $T_n$,
rather than $T_n^{(1)}$, for practical use.

{\bf Study 4.}
In this study, a nonlinear single-index null model is considered.
We try four alternatives with different structural dimension as follows:
\benrr
&&H_{16}:\, \, Y=(\b^{\top}X)^3+a|\b^{\top}X|+\ep\cr
&&H_{17}:\, \, Y=(\b^{\top}X)^3+a X_3^2+\ep\cr
&&H_{18}:\, \, Y=(\b^{\top}X)^3+a(X_2/4+|X_3^2|+\cos(\pi X4))+\ep\cr
&&H_{19}:\, \, Y=(\b^{\top}X)^3+a(X_2/2+X_3^2+\cos(\pi X_4)+X_5\exp(X_6/2)+X_8X_7)+\ep\cr
\eenrr
Let $p=4$ for $H_{16}$, $H_{17}$, $H_{18}$ and $p=8$ for $H_{19}$. $\b=[1,0,\cdots,0]^{\top}$.
$\Si=\Si_1$, $\si_u=0.5$.
$a$ is designed to be $0, 0.2, 0.4, 0.6, 0.8, 1.0$.
In these cases, $q$ is always 1 for the null but different for alternatives.
For $H_{16}$, $q=1$ for any nonzero $a$.
The structure dimension under  $H_{17}$ is 2, and under $H_{18},$ $p=q=4$.
For $H_{19}$, $p=q=8$.
The test $T_n$ uses the same bandwidths as chosen for linear model
above. For $T_n^{Zh}$, we adjust bandwidths to keep its performance.
Set $c=2.7$ for $H_{16}$, $H_{17}$, $H_{18}$ and $c=3$ for $H_{19}$.
The results are presented in Figure 2.
\begin{center}
Figure~2. about here
\end{center}

We have the following observations.
First, the model-adaptive method $T_n$ has greater empirical power than $T_n^{Zh}$
for  all chosen alternatives. Under $H_{18}$ and $H_{19}$, though
convergence rate of the two teats are same, $T_n$ is still more  powerful
than $T_n^{Zh}$. Because $T_n$ is constructed by
$nh^{1/2}V_n/\sqrt{\Si}=h^{(1-q/2)}\times nh^{q/2}V_n/\sqrt{\Si}$.
Secondly, the power of $T_n^{Zh}$ decreases quickly as $p$ increases
while that of $T_n$ does not.

\clearpage

\section{Appendix. Proofs}

This section is organized as follows.
In Section 5.1,  Proposition~\ref{pro2} is proved. The proof of Theorem~\ref{th3.4}
appears in Section~ 5.2. Based on the  asymptotic behavior of $\h \b$ and $\h B$
under the local alternatives,  the proof of Theorem~\ref{th3.5} is included in
Section~5.3.  As Theorem~\ref{theo3.2} is a special case of Theorem~\ref{th3.5} when
$C_n=0$, its proof is omitted. In Section 5.4, we only sketch
the proof of Theorem~\ref{theo3.1} as it is similar to that of Theorem~\ref{th3.5}.
Section 5.5 shows a sketch of the proof for Theorem~\ref{theo3.3}.

\subsection{Proof of Proposition ~\ref{pro2}}\lel{prop2}
The claim (1) has been proved in Lee and Sepanski (1995).
We now prove the claim (2).
Recall some notation: ${\mathbf X}$ is  $n\times p$ matrix whose $i$th row is
$x_i^{\top}, i=1,\cdots, n$, ${\mathbf X}_v$ is the $N\times p$ matrix whose $s$th row
is $\ti x_s^{\top}, s=1,\cdots, N$, and ${\mathbf Y}$ is a $n\times 1$ vector, while
$g({\mathbf X}_v\b)$ represents the $N\times 1$ vector and equals to $[g(\b^{\top}\ti
x_{1}),\cdots,g(\b^{\top} \ti x_N)]^{\top}$.
The matrix ${\mathbf D}$ is the $n\times k$ matrix whose $i$-th row $\bar w^{\top}_i$
is a $1\times k$ vector consist of polynomials of $w_i$.
The matrix ${\mathbf D}_v$ is the corresponding matrix of validation data, whose $s$-th
row  $\bar w^{\top}_s$ is a vector consist of polynomials of $\ti w_s$.
For linear model, $\bar w_i= w_i$ and $\bar w_s=\ti w_s$.
For nonlinear model, we let $\bar w_i$ be a vector consisting of a constant and the
first two order polynomials of $ w_i$.

Let
\benrr
Q_{n}(\b)=\frac{1}{n}\Big({\mathbf Y}-{\mathbf D}({\mathbf D}_v^{\top}{\mathbf D}_v)^{-1}
{\mathbf D}_v^{\top}g({\mathbf X}_v\b)\Big)^{\top}
\Big({\mathbf Y}-{\mathbf D}({\mathbf D}_v^{\top}{\mathbf D}_v)^{-1}{\mathbf D}_v^{\top}
g({\mathbf X}_v\b)\Big).
\eenrr
The estimator $\h \b$ satisfies the first order condition: $\partial Q_n(\h \b)/\partial \b=0$.
By Taylor expansion and the mean value theorem:
\benrr
&&\Big[\frac{\partial g^{\top}({\mathbf X}_v\b_0)}{\partial\b}{\mathbf D}_v\Big]
({\mathbf D}_v^{\top}
{\mathbf D}_v)^{-1}{\mathbf D}^{\top}({\mathbf Y}-{\mathbf D}({\mathbf D}_v^{\top}
{\mathbf D}_v)^{-1}{\mathbf D}_v^{\top}g({\mathbf X}_v\b_0))\\
&=&\Big \{\Big[\frac{\partial^2 g^{\top}({\mathbf X}_v\bar{\b})}{\partial \b\partial\b^{\top}}{\mathbf D}_v\Big]({\mathbf D}_v^{\top}{\mathbf D}_v)^{-1}
{\mathbf D}^{\top}({\mathbf Y}-{\mathbf D}({\mathbf D}_v^{\top}{\mathbf D}_v)^{-1}{\mathbf D}_v^{\top}
g({\mathbf X}_v\bar{\b}))\\
 &&-\Big[\frac{\partial g^{\top}({\mathbf X}_v\bar{\b})}{\partial\b}{\mathbf D}_v\Big]({\mathbf D}_v^{\top}
 {\mathbf D}_{v})^{-1}({\mathbf D}^{\top}{\mathbf D})({\mathbf D}_v^{\top}{\mathbf D}_{v})^{-1}[\frac{\partial g^{\top}({\mathbf X}_v\bar{\b})}{\partial\b}{\mathbf D}_v]\Big\}(\b_0-\h \b)
\eenrr
where $\bar \b$ is a vector satisfying $\|\bar \b-\b\|\le \|\h \b-\b_0\|$, and
\benrr
[\frac{\partial^2 g^{\top}({\mathbf X}_v\bar{\b})}{\partial \b\partial\b^{\top}}{\mathbf D}_v]=[\frac{\partial^2 g^{\top}({\mathbf X}_v\bar{\b})}{\partial \b\partial\b_1}{\mathbf D}_v, \cdots, \frac{\partial^2 g^{\top}({\mathbf X}_v\bar{\b})}{\partial \b\partial\b_p}{\mathbf D}_v].
\eenrr
Let $g'$, $g''$ denote the first and second derivatives of $g$, respectively. By the LLNs,
\benrr
&&\frac{1}{N}\frac{\partial g^{\top}({\mathbf X}_v\b)}{\partial\b}{\mathbf D}_v
=\frac{1}{N}\sum_{s=1}^{N}g{'}(\b^{\top}\ti x_s)\ti x_s\bar w_s^{\top}\to_p  E[g{'}(\b^{\top}X)X\bar W^{\top}],\\
&& \frac{1}{N}\frac{\partial^2 g^{\top}({\mathbf X}_v\bar{\b})}{\partial
\b\partial\b_l}{\mathbf D}_v \to_p E[ g{''}(\b^{\top}X) X_{(l)}
X\bar W^{\top}],
\eenrr
and
\benrr
\lefteqn{
\frac{1}{n} {\mathbf D}^{\top}( Y-{\mathbf D} ({\mathbf D}_v^{\top}{\mathbf D}_v)^{-1}
{\mathbf D}_v^{\top}g({\mathbf X}_v\bar{\b}) ) }\\
&=& C_n E[\bar W G(X)]+(E[\bar W g(\b_0^{\top}X)]-E(\bar W\bar W^{\top})\g_0)+o_p(1)\\
&=& o_p(1),
\eenrr
where $\g_0=E^{-1}(\bar W\bar W^{\top})E[\bar W g(\b_0^{\top}X)]$. Hence 
\benrr
\h \b-\b_0 &=& \bigg\{[\frac{\partial^2 g^{\top}({\mathbf X}_v\bar{\b})}{\partial \b\partial\b^{\top}}{\mathbf D}_v]({\mathbf D}_v^{\top}{\mathbf D}_v)^{-1}{\mathbf D}^{\top}({\mathbf Y}-{\mathbf D}({\mathbf D}_v^{\top}{\mathbf D}_v)^{-1}{\mathbf D}_v^{\top}g({\mathbf X}_v\bar{\b}))\\
&&-[\frac{\partial g^{\top}({\mathbf X}_v\bar{\b})}{\partial\b}{\mathbf D}_v]({\mathbf D}_v^{\top}{\mathbf D}_{v})^{-1}({\mathbf D}^{\top}{\mathbf D})({\mathbf D}_v^{\top}{\mathbf D}_{v})^{-1}[\frac{\partial g^{\top}({\mathbf X}_v\bar{\b})}{\partial\b}{\mathbf D}_v]\bigg\}^{-1}\\
&&\times [\frac{\partial g^{\top}({\mathbf X}_v\b_0)}{\partial\b}{\mathbf D}_v]({\mathbf D}_v^{\top}{\mathbf D}_v)^{-1}{\mathbf D}^{\top}({\mathbf Y}-{\mathbf D}({\mathbf D}_v^{\top}{\mathbf D}_v)^{-1}{\mathbf D}_v^{\top}g({\mathbf X}_v\b_0))\cr
&=&\left\{E[g{'}(\b^{\top}X)X\bar W^{\top}]E^{-1}[\bar W\bar W^{\top}]E[g{'}(\b^{\top}X)\bar W X^{\top}]+O_p(C_n)\right\}^{-1}\\
&&\times \left\{E[g{'}(\b^{\top}X)X\bar W^{\top}]E^{-1}[\bar W\bar W^{\top}]\right\}\frac{1}{n}{\mathbf D}^{\top}({\mathbf Y}-{\mathbf D}({\mathbf D}_v^{\top}{\mathbf D}_v)^{-1}{\mathbf D}_v^{\top}g({\mathbf X}_v\b)).
\eenrr
On the other hand,
\benrr
\lefteqn{
\frac{1}{n}{\mathbf D}^{\top}({\mathbf Y}-{\mathbf D}({\mathbf D}_v^{\top}{\mathbf D}_v)^{-1}{\mathbf D}_v^{\top}g({\mathbf X}_v\b)) }\\
&=&\frac{1}{n}{\mathbf D}^{\top}C_n G({\mathbf X})+\frac{1}{n}{\mathbf D}^{\top}(g({\mathbf X}\b)+\vep-{\mathbf D}({\mathbf D}_v^{\top}{\mathbf D}_v)^{-1}{\mathbf D}_v^{\top}g({\mathbf X}_v\b))\\
&=&\frac{C_n}{n}\sum_{i=1}^n\ti w_i G(x_i)+\frac{1}{n}{\mathbf D}^{\top}(g({\mathbf X}\b)+\vep-{\mathbf D}E^{-1}[\bar W\bar W^{\top}]E[\bar W^{\top} g(\b^{\top} X)])\\
 &&-\left(\frac{1}{n}{\mathbf D}^{\top}{\mathbf D}\right)\left[\frac{1}{N}{\mathbf D}_v^{\top}{\mathbf D}_v\right]^{-1}\frac{1}{N}({\mathbf D}_v^{\top}g({\mathbf X}_v\b)-{\mathbf D}_v^{\top}{\mathbf D}_vE^{-1}[\bar W^{\top}\bar W]E[\bar W^{\top}g(\b^{\top}X)])\\
&=&C_n E[\bar WG(x)]+O_p(1/\sqrt{n})+O_{p}(1/\sqrt{N}).
\eenrr
This completes the proof of part (2) of Proposition \r{pro2}.
 \hfill $\fbox{}$

\subsection{Proof of Theorem~\ref{th3.4}}

Denote $\z=\Cov(X,W)\Si_W^{-1}W$.
In the discretization step, we construct new samples $(\z_i, I(y_i\leq y_j))$.
For each $y_j$, we estimate $\L(y_j)$ which spans $S_{I(Y\leq y_j)|\z}$ by using SIR and denote the estimate by $\L_n(y_j)$.
In the expectation step, we estimate $\L=E[\L(t)]$, which spans $S_{Y|\z}$, by $\L_{n,n}=n^{-1}\sum_{j=1}^{n}\L_n(y_j)$.
Let $\la_1> \la_2> \cdots > \la_q> \la_{q+1}=0=\cdots = \la_p$ be the descending sequence of eigenvalues of the matrix  $\L$ and
$\h \la_1\geq\h \la_2\geq\cdots\geq\h \la_p$ be the descending sequence of eigenvalues of the matrix  $\L_{n,n}$.
Recall the $D_n$ in $\h q$ of (\r{qbic}) was selected as $\sqrt{n}$.
Define the objective function in (\ref{qbic}) as
\benrr
G(l)=\frac{n}{2}\times\frac{\sum_{i=1}^{l} \{\log(\h \la_i+1)-\h \la_i\}}{\sum_{i=1}^{p} \{\log(\h \la_i+1)-\h \la_i\}}-2\times n^{1/2}\times \frac{l(l+1)}{2p}.
\eenrr
Now we prove that for any $l>1$, $P(G(1)>G(l))\to 1$, i.e.,  $P(\h q=1)\to 1$.
\benrr
G(1)-G(l)= n^{1/2} \times \frac{l(l+1)-2}{p}-\frac{n}{2}\times\frac{\sum_{i=2}^{l} \{\log(\h \la_i+1)-\h \la_i\}}{\sum_{i=1}^{p} \{\log(\h \la_i+1)-\h \la_i\}}
\eenrr
If $\L_{n,n}-\L=O_{p}(C_{n})$, then $\h \la_i-\la_i=O_{p}(C_{n})$.
By the second order Taylor Expansion, we have
$\log(\h \la_i+1)-\h \la_i=-\h \la^2_i+o_{p}(\h \la^2_i)$.
Thus, $\sum_{i=2}^{l} \{\log(\h \la_i+1)-\h \la_i\}=O_{p}(C^{2}_{n})$ and $\sum_{i=1}^{p} \{\log(\h \la_i+1)-\h \la_{i}\}$ converge to a negative constant in probability.
Since $nC_{n}^{2}/n^{1/2}\to 0$ and $l(l+1)>2$, $P(G(1)>G(l))\to 1$.

Now we check  the condition of  $\L_{n,n}-\L=O_{p}(C_{n})$.
First, we investigate the convergence rate of $\L_{n}(t)-\L(t)$ for any fixed $t$.
We  have
\benrr
\L(t)=\Si_{\z}^{-1}\Var(E[\z|\ti Y(t)])p(1-p)=\Si_X^{-1}\Si_W\Si_X^{-1}\Var(E[\z|\ti Y(t)])p(1-p).
\eenrr
It is easy to see that
\benrr
\Var(E[\z|\ti Y(t)])=(u_{1}-u_{0})(u_{1}-u_{0})^{\top}p(1-p)
\eenrr
where $p=P(Y\leq t)=E(I(Y\leq t))$, $u_{i}=E[\z|\ti Y(t)=i]$, $i=0,1$. Further, $u_{1}-u_{0}$ can be rewritten as
\benrr
u_{1}-u_{0}=\left\{E[\z I(Y\leq t)]-E[\z]E[I(Y\leq t)]\right\}/(p(1-p)).
\eenrr
We can use the matrix
\benrr
\L(t)=\Si_X^{-1}\Si_W\Si_X^{-1}\left[E\{(\z-E(\z))I(Y\leq t)\}\right]\left[E\{(\z-E(\z))I(Y\leq t)\}\right]^{\top}
\eenrr
to identify the central subspace we want.
Denote $m(t)=E[(\z-E(\z))I(Y\leq t)].$
The sample version of $m(t)$ is
\benrr
\h m(t)=\frac{1}{n}\sum_{i=1}^{n}(\z_i-\bar \z)I(y_i\leq t),
\eenrr
where $\z_i=\h \Cov(X,W)\h \Si_W^{-1}w_i$ and $\bar \z=(1/n)\sum_{i=1}^n \z_i$.
Let $Y_{a}$ be the response under the local alternative, then
\benrr
\h m(t)-m(t)&=&\frac{1}{n}\sum_{i=1}^{n}(\z_{i}-\bar{\z})I(y_{i}\leq t)-E\{(\z-E(\z))I(Y\leq t)\}\\
            &=&\frac{1}{n}\sum_{i=1}^{n}(\z_{i}-\bar{\z})I(y_{i}\leq t)-E\{(\z-E(\z))I(Y_{a}\leq t)\}\\
            &&+E\{(\z-E(\z))I(Y_{a}\leq t)\}-E\{(\z-E(\z))I(Y\leq t)\}.
\eenrr
The convergence rate of the first term in the right hand side is $O_{p}(\sqrt{n})$. For simplicity, we assume $E(\z)=0$. The second term is
\benrr
E[\z I(Y_{a}\leq t)]-E[\z I(Y\leq t)]&=&E\left\{\z[P(Y_{a}\leq t
|\z)-P(Y\leq t |\z)] \right\}
\eenrr
Since $\z=\Si_{X}\Si_{W}^{-1}W$,
\benrr
\lefteqn{
P(Y_{a}\leq t |\z)-P(Y\leq t |\z)} \\
&=&P(Y_{a}\leq t |W)-P(Y\leq t |W)=F_{Y|W}(t-C_{n}E[G(B^{\top}X)|B^{\top}W])-F_{Y|W}(t)\\
&=&-C_{n}E[G(B^{\top}X)|B^{\top}W]f_{Y|W}(t)+O_{p}(C_{n}^{2}).
\eenrr
Thus, we have $E\{(\z-E(\z))I(Y_{a}\leq t)\}-E\{(\z-E(\z))I(Y\leq t)\}=O_{p}(C_{n})$.
Altogether,  $\L_{n}(t)-\L(t)=O_{p}(C_{n})$, for each $t\in \R$.
Finally, similar to the proof for Theorem 3.2 of Li et al.\,\,(2008) the condition $\L_{n,n}-\L=O_{p}(C_{n})$ holds. \hfill $\fbox{}$

\subsection{Proof of Theorem~\ref{th3.5}}
In this subsection, we first prove (ii)  which is the large sample property of $V_n$ under the local alternatives and then give a sketch of the proof of (i).
For the local alternatives in (\ref{localalterB}), according to Theorem~\ref{th3.4}, $\h q=1$ with a probability going to  $1$.
Thus, we can only work on the event that $\h q=1$.  Note that $\h B(\h q)$ converges to $\b/\|\b\|$ in probability rather than the $p \times q$ matrix $B$ that is the dimension reduction base matrix of the central mean subspace.
In other words,  $\h B$ is not a consistent estimate of $B$.
However, in this proof, we still use $B$ to write  the limit of $\h B$ for notation simplicity.
By Proposition~\ref{pro2}, we have
\benr\label{localbeta}
\h \b-\b=&C_n H(\b)(1+o_p(1)).
\eenr
where
\benrr
H(\b)&=&\left\{E[g{'}(\b^{\top}X)X\bar W^{\top}]E^{-1}[\bar W\bar W^{\top}]E[g{'}(\b^{\top}X)\bar W X^{\top}]\right\}^{-1}\cr
     &&\times E[g{'}(\b^{\top}X)X\bar W^{\top}]E^{-1}[\bar W\bar W^{\top}]E[\bar W G(B^{\top}X)].
\eenrr
Let $G_i=G(z_i)$ and $\D_i=\D(z_i)$, where $z_i=B^{\top}w_i$, $G$ is as
in (\ref{localalterB}), and $\D$ as in  (\ref{defs1}).
Recall the notation from (\ref{rhat}) and (\ref{defs2}). Rewrite
\benrr
\h e_i= g_i+C_n G_i+ \vep_i-\h r_i = r_i-\h r_i+C_n G_i+ e_i.
\eenrr

Recalling  $\h z_i=\h B^{\top}w_i$,
we obtain the following decomposition for $V_{n}$.

\benr\label{decom1}
V_{n}&=&\frac{1}{n(n-1)}\sum_{i=1}^{n}\sum_{j\neq i}^{n}K_{h}(\h z_i-\h z_j)(e_{i}+C_{n}G_i)(e_{j}+C_{n}G_j)\\
      &&+\frac{1}{n(n-1)}\sum_{i=1}^{n}\sum_{j\neq i}^{n}K_{h}(\h z_i-\h z_j)(e_{i}+C_{n}G_i)(r_{j}-\h r_{j(2)}) \nn \\
      &&+\frac{1}{n(n-1)}\sum_{i=1}^{n}\sum_{j\neq i}^{n}K_{h}(\h z_i-\h z_j)(r_{i}-\h r_{i(1)})(e_{j}+C_{n}G_j) \nn  \\
      &&+\frac{1}{n(n-1)}\sum_{i=1}^{n}\sum_{j\neq i}^{n}K_{h}(\h z_i-\h z_j)(r_{i}-\h r_{i(1)})(r_{j}-\h r_{j(2)}) \nn \\
    &=:&V_{n1}+V_{n2}+V_{n3}+V_{n4}, \qquad \mb{say}. \nn
\eenr
We now deal with  $V_{ni}$'s in the following steps.
\begin{step}\label{step5.1}
$nh^{1/2}V_{n1}\to_DN( \nu_1 , \t_1)$,
where $\t_1$ is as in (\r{si123}) and
\benr\lel{nu1}
\nu_1=E[\D^2(Z)f(Z)].
\eenr
\end{step}
{\it Proof:}
It follows from (\ref{decom1}) that
\benr\label{v1}
V_{n1}&=&\frac{1}{n(n-1)}\sum_{i=1}^{n}\sum_{j\neq i}^{n}K_{h}(\h z_i- \h z_j)e_{i}e_{j}+2C_{n}\frac{1}{n(n-1)}\sum_{i=1}^{n}\sum_{j\neq i}^{n}K_{h}(\h z_i-\h z_j)e_{i}G_j  \\
      & &+C_{n}^{2}\frac{1}{n(n-1)}\sum_{i=1}^{n}\sum_{j\neq i}^{n}K_{h}(\h z_i -\h z_j)G_iG_j \nn\\
      &=:& I_{1}+2C_{n}I_{2}+C_{n}^{2}I_{3}. \nn
\eenr

{\it Step 5.1.1.} Deal with $I_1 $. Rewrite $I_1= I_{1,1}+I_{1,2}$, where
\benrr
I_{1,1}&=&\frac{1}{n(n-1)}\sum_{i=1}^{n}\sum_{j\neq i}^{n}K_{h}(z_i-z_j)e_{i}e_{j},\\
I_{1,2}&=& \frac{1}{n(n-1)}\sum_{i=1}^{n}\sum_{j\neq i}^{n}(K_{h}(\h z_i- \h z_j)-K_{h}(z_i-z_j))e_{i}e_{j}.
\eenrr
Following Lemma 3.3a of Zheng (1996)
we obtain $nh^{1/2}I_{1,1}\to_D  N(0, \t_1)$, where
\benrr
\t_1=2\int(\si^{2}(z))^2f^2(z)dz\int K^{2}(u)du.
\eenrr
The Taylor expansion yields that
\benrr
I_{1,2}=\frac{(\h B-B)^{\top}}{h}\frac{1}{n(n-1)}\sum_{i=1}^{n}\sum_{j\neq i}^{n}K{'}(\frac{z_i-z_j}{h})\frac{w_{i}-w_{j}}{h}e_{i}e_{j}(1+o_p(1)).
\eenrr
Let
\benrr
I_{1,2}^{*}=\frac{1}{(n-1)n}\sum_{i=1}^{n}\sum_{j\neq i}^{n}K{'}(\frac{z_i-z_j}{h})\frac{w_i-w_j}{h}e_i e_j.
\eenrr
Similarly as $I_{1,1}$, $I_{1,2}^{*}$ is a degenerate U-statistic with kernel
\benrr
H_{n}((y_{i},w_{i}),(y_{j},w_{j}))=K{'}(\frac{z_i-z_j}{h})\frac{w_{i}-w_{j}}{h}e_{i}e_{j}.
\eenrr
Combining $\|\h B-B\|_{2}=O_{p}(C_n)$ and $nh^{5/2}\to \iny$, we obtain $nh^{1/2}I_{12}=o_{p}(1)$. Hence $nh^{1/2}I_{1}\to_D  N(0, \t_1)$.
\vs .1cm
{\it Step 5.1.2}. Next, consider $I_2$. Rewrite $I_2= I_{2,1}+I_{2,2},$ where
\benrr
I_{2,1}&=&\frac{1}{n(n-1)}\sum_{i=1}^{n}\sum_{j\neq i}^{n}K_{h}(z_i-z_j)e_i G_j, \\
I_{22}&=&\frac{1}{n(n-1)}\sum_{i=1}^{n}\sum_{j\neq i}^{n}(K_ {h}(\h z_i - \h
z_j)-K_h(z_i-z_j))e_iG_j.
\eenrr
By computing the second order moment, we know $I_{2,1}=O_p(1/\sqrt{n})$. As to $I_{2,2}$,
\benrr
I_{2,2}=\frac{\h B-B}{h}\frac{1}{n(n-1)}\sum_{i=1}^{n}\sum_{j\neq i}^{n} K{'} (\frac{ z_i - z_j} {h} )\frac{ w_i-w_j }{ h } e_i G_j (1+o_p(1)).
\eenrr
Let
\benrr
I_{2,2}^{*}=\frac{1}{n(n-1)}\sum_{i=1}^{n}\sum_{j \neq
i}^{n} K{'} (\frac{z_i-z_j}{h})\frac{w_i-w_j}{h}e_i G_j.
\eenrr
Since the kernel function $K(\cdot)$ is symmetric, $I_{2,2}^{*}$ can be rewritten as a non-degenerate U-statistic.
Thus $I_{2,2}^{*}=O_p(1/\sqrt{n})$.
Combining the convergence rates of $I_{2,1}$ and $I_{2,2}$, we know that
$nh^{1/2}C_nI_2=o_p(1)$.
\vs .1cm
{\it Step 5.1.3.}   Consider $I_3$. It is easy to see that  $I_3\to_p  E[\D^2(Z)f(Z)],$ where $Z=B^{\top}W$.
\vs .1cm
Summarizing the above results for $I_1$, $I_2$ and $I_3$, we have that  if  $C_{n}=n^{-1/2}h^{-1/4}$,
$
nh^{1/2}V_{n1}\to_D N(\nu_1 , \t_{1}),
$
thereby completing the proof of Step 5.1. \hfill $\fbox{}$

\begin{step}\label{step5.2}
$
nh^{1/2}V_{n2}\to_DN\left(\nu_2, 2\la^{-1}\t_2\right),
$
where $\t_2$ is defined in (\r{si123}) and
\benr\lel{nu2}
\nu_2=-E\{\D(Z)E[g{'}(\b^{\top}X)X^{\top}|Z]f(Z)\}H(\b_0).
\eenr
\end{step}
{\it Proof:}
Rewrite $V_{n2}$ as
\benr\label{V12}
V_{n2}&=&\frac{1}{n(n-1)}\sum_{i=1}^{n}\sum_{j\neq i}^{n}K_h (\h z_i-\h z_j)e_{i}(r_{j}-\h r_{j(2)})\\
       &&+\frac{C_{n}}{n(n-1)}\sum_{i=1}^{n}\sum_{j\neq i}^{n}K_h (\h z_i- \h z_j)G_i(r_{j}-\h r_{j(2)}) \nn \\
      &=:&V_{n2,1}+C_{n}V_{n2,2}, \qquad \mb{say}.  \nn
\eenr
{\it Step 5.2.1.} Deal with the term $V_{n2,1}$. It  can be decomposed as
\benrr
V_{n2,1}&=&\frac{1}{n(n-1)}\sum_{i=1}^{n}\sum_{j\neq i}^{n}K_{h}(z_i-z_j)e_{i}(r_{j}-\h r_{j(2)})\\
         &&+\frac{1}{n(n-1)}\sum_{i=1}^{n}\sum_{j\neq i}^{n}(K_h(\h z_i-\h z_j)-K_{h}(z_i-z_j))e_{i}(r_{j}-\h r_{j(2)}).
\eenrr
Recalling the definition of the  estimator of $r_{(2)}(\b^{\top}w,\b)$ in (\r{rhat}), we have
\begin{equation}\label{decomr}
\begin{split}
r_{j}-\h r_{j(2)} =&\frac{2}{N}\sum_{s=N/2+1}^{N}M_{v_N}(\h \b^{\top}w_j-\h \b^{\top}\ti w_s)(r_{j}-\h{\ti g}_{s})/\frac{2}{N} \sum_{s=N/2+1}^{N} M_{v_N}(\h \b^{\top}w_j-\h \b^{\top}\ti w_s),
\end{split}
\end{equation}
where $\h{\ti g}_{s}$ is defined in (\ref{defs2}).
In order to analyze $r_{j}-\h r_{j(2)}$ further, we need the following entities. Let
\benr\label{deffN2}
&& \bar f_{N(2)}(x)=\frac{2}{N}\sum_{s=N/2+1}^{N}M_{v_N}(x-\b^{\top}\ti w_s),\,\, \h{\bar f}_{N(2)}(x) = \frac{2}{N} \sum_{s=N/2+1}^{N} M_{v_N}(x-\h \b^{\top}\ti w_s), \\
\lel{Q123(2)}
&&Q_{1(2)}(\b^{\top}w_j)=\frac{2}{N}\sum_{s=N/2+1}^{N}M_{v_N}(\b^{\top}w_j-\b^{\top}\ti w_s)(r_j-\ti r_s),\\ &&Q_{2(2)}(\b^{\top}w_j)=\frac{2}{N}\sum_{s=N/2+1}^{N}M_{v_N}(\b^{\top}w_j-\b^{\top}\ti w_s)(\ti r_s-\ti g_s),\nn \\
&&Q_{3(2)}(\b^{\top}w_j)=\frac{2}{N}\sum_{s=N/2+1}^{N}M_{v_N}(\b^{\top}w_j-\b^{\top}\ti w_s)(\ti g_s-\h{\ti g}_s). \nn
\eenr
The kernel function $M_{v_N}(\h \b^{\top}w_j-\h \b^{\top}\ti w_s)$ in the numerator of (\ref{decomr}) can be rewritten as
$$M_{v_N}(\b^{\top}w_j-\b^{\top}w_s)+[M_{v_N}(\h \b^{\top}w_j-\h \b^{\top}w_s)-M_{v_N}(\b^{\top}w_j-\b^{\top}w_s)],$$
and the denominator can be decomposed as
$$\frac{1}{\bar f_{N(2)}(\b^{\top}w_j)}+[\frac{1}{\h{\bar f}_{N(2)}(\h \b^{\top}w_j)}-\frac{1}{\bar f_{N(2)}(\b^{\top}w_j)}].$$
Further,
write
\benrr
r_{j}-\h{\ti g}_{s}= [r_{j}-\ti r_{s}]+[\ti r_s-\ti g_s]+[\ti g_s-\h{\ti g}_s].
\eenrr
Combining the above decompositions into (\ref{decomr}), $r_{j}-\h r_{j(2)}$ can be decomposed into 12 terms, and
then $V_{n2,1}$ can  be decomposed into 24 terms.
We only consider the following three terms that make non-negligible
contribution. The remaining terms can be shown to be asymptotically negligible,
in probability. Accordingly, consider
\benr\lel{I456}
I_4 &=& \frac{1}{n(n-1)}\sum_{i=1}^{n}\sum_{j\neq i}^{n}K_{h}(z_i-z_j)e_{i}Q_{1(2)}(\b^{\top}w_j)/\bar f_{N(2)}(\b^{\top}w_j), \\
I_5 &=& \frac{1}{n(n-1)}\sum_{i=1}^{n}\sum_{j\neq i}^{n}K_{h}(z_i-z_j)e_{i}Q_{2(2)}(\b^{\top}w_j)/\bar f_{N(2)}(\b^{\top}w_j), \nn \\
I_6 &=& \frac{1}{n(n-1)}\sum_{i=1}^{n}\sum_{j\neq i}^{n}K_{h}(z_i-z_j)e_{i}Q_{3(2)}(\b^{\top}w_j)/\bar f_{N(2)}(\b^{\top}w_j) \nn
\eenr
where $\bar f_{N(2)}(\b^{\top}w_j)$ is defined in (\ref{deffN2}), and $Q_{1(2)}(\cdot)$, $Q_{2(2)}(\cdot)$, $Q_{3(2)}(\cdot)$ are in (\ref{Q123(2)}).
Let $\bar f$ denote the density of $\b^{\top}W$.

We first prove that $nh^{1/2}I_{4}=o_{p}(1)$. Rewrite $I_4=n^{-1}\stj I_{41}(z_j)\times I_{42}(\b^{\top}w_j)$, where
\benrr
I_{41}(z_j)=\frac{1}{(n-1)}\sum_{i\neq j}^{n}K_{h}(z_i-z_j)e_{i},\quad I_{42}(\b^{\top}w_j)= \frac{Q_{1(2)}(\b^{\top}w_j)}{\bar f_{N(2)}(\b^{\top}w_j)}.
\eenrr
Thus, the application of Cauchy - Schwarz inequality yields that $|I_4|\leq \sqrt{(1/n)\stj I^2_{41}(z_j)}\times \sqrt{(1/n)\stj I^2_{42}(\b^{\top}w_j)}$. We only need to  bound the conditional expectations $E[I^2_{41}(z_j)]$ and $E[I^2_{42}(\b^{\top}w_j)]$ when $z_j$, $\b^{\top}w_j$ are given.
For $I_{41}(z_j)$,
\benrr
E[I^2_{41}(z_j)]=&\frac{1}{(n-1)^2}E[(\sum_{i\neq j}^{n}K_{h}(z_i-z_j)e_{i})^2]=\frac{1}{(n-1)h^2}E[K^2(\frac{z_i-z_j}{h})e^2_{i}]= O(\frac{1}{nh}).
\eenrr
For $I_{42}$, we can obtain that given $\b^{\top}w_j$,
\benrr
|I_{42}(\b^{\top}w_j)|\leq \left|\frac{Q_{1(2)}(\b^{\top}w_j)}{\bar f(\b^{\top}w_j)} \right|\sup_{\b^{\top}w_j}\left| \frac{\bar f(\b^{\top}w_j)}{\bar f_{N(2)}(\b^{\top}w_j)}\right|.
\eenrr
Since
\benrr
\sup_{\b^{\top}w_j}|\bar f_{N(2)}(\b^{\top}w_j)-\bar f(\b^{\top}w_j)|=o_p(1),\,\, \sup_{\b^{\top}w_j}\left| \frac{\bar f_{N(2)}(\b^{\top}w_j)}{\bar f(\b^{\top}w_j)}-1\right|=o_p(1),
\eenrr
and $\bar f_(\b^{\top}w_j)$ is uniformly bounded below, we only need to bound $Q_{1(2)}^2(\b^{\top}w_j)$ in the numerators.
But
\begin{equation*}
\begin{split}
E[Q_{1(2)}^2(\b^{\top}w_j)]=&\frac{N(N-2)}{N^2v_N^2}E[M(\frac{\b^{\top}w_j-\b^{\top}\ti w_s}{v_N})(r_j-\ti r_s)M(\frac{\b^{\top}w_j-\b^{\top}\ti w_{s^{'}}}{v_N})(r_j-\ti r_{s^{'}})]\cr
                          &+\frac{2}{Nv_N^2}E[M^2(\frac{\b^{\top}w_j-\b^{\top}\ti w_s}{v_N})(r_j-\ti r_s)^2]\cr
                      \leq&C_1 v_N^{2\ell}+N^{-1}C_2v_N,
\end{split}
\end{equation*}
where $C_1$ and $C_2$ are two constants.
The last inequality is obtained by Conditions (f),(r) and (M). Thus
$E[I^2_{42}(\b^{\top}w_j)]$ is bounded from the above by $C_1
v_N^{2\ell}+C_2v_N/N$, in probability.
Summarizing the results of $E[I^2_{41}]$ and $E[I^2_{42}]$, we have  $|nh^{1/2}I_4|\leq nh^{1/2}O_p(\frac{1}{\sqrt{nh}}\sqrt{v_N^{2\ell}+\frac{v_N}{N}})=o_p(1)$.
\vs .1cm

Consider $I_{5}$.  Rewrite it as $I_{5}=I_{51}+I_{52}$, where
\benr\label{I5}
I_{51}=E[I_{5}|\ti \eta_s, \ti z_s ,z_i,e_i], \quad I_{52}=(I_{5}-E[I_{5}|\ti \eta_s, \ti z_s ,z_i,e_i]).
\eenr
Note that
\benrr
I_{51}&=&\frac{2}{nN}\sum_{j=1}^{n}\sum_{s=N/2+1}^{N}e_i\ti \eta_s \int\frac{1}{h}K(\frac {z_i-z_j}{h}) \frac{1}{v_N}M(\frac{\b^{\top}w_j-\b^{\top}\ti w_s}{v_N})d(\b^{\top}w_j)\cr
&=&\frac{2}{nN}\sum_{j=1}^{n}\sum_{s=N/2+1}^{N}e_i\ti \eta_s\int\frac{1}{h}K(\frac{z_i-\ti z_s-v_Nu/\|\b\|}{h})\frac{1}{v_N}M(u)d(\b^{\top}\ti w_s+v_Nu).
\eenrr
The second equation holds because $z_j=B^{\top}w_j=\b^{\top}w_j/\|\b\|$. Further,
\benrr
\int\frac{1}{h}K(\frac{z_i-\ti z_s -v_Nu/\|\b\|}{h})M(u)du =\frac{1}{h}K(\frac{z_i-\ti z_s}{h}) +\frac{1}{h}K^{''}(\frac{z_i-\ti z_s}{h})\frac{v_N^2\|\b\|^2}{h^2}.
\eenrr
Thus, $I_{51}=\frac{2}{nN}\sum_{i=1}^{n}\sum_{s=N/2+1}^{N}e_{i}\ti \eta_s K_h(z_i-\ti z_s)(1+o_p(1)).$
By Central Limit Theorem we have
\benrr
\sqrt{\frac{nN}{2}}h^{1/2}I_{5,1}\to_D N(0, \int K^{2}(u)du\int \si^{2}(z)\xi^{2}(z)f^{2}(z)dz),
\eenrr
where $\si^2(Z)$ and $\xi^2(Z)$ are defined in (\ref{defs1}).
By some elementary calculations, we can derive that $E[(I_{52})^{2}]=O_{p}(1/(n^2Nhv_N)).$
 Chebyshev's inequality yields that $nh^{1/2}I_{52}=o_{p}(1)$. Hence
\benr
nh^{1/2}I_{5}\to_D N\Big (0, 2\la^{-1}\int K^{2}(u)du\int \si^{2}(z)\xi^{2}(z)f^{2}(z)dz \Big ).
\eenr

Now consider $I_6$. Recall the definition of $Q_{3(2)}$ in (\ref{Q123(2)}) and the
definition of $\ti g$ below  (\ref{defs2}). Taylor expansion
of the function $\ti g$ yields that
$I_6=I_6^{*}(\b-\h \b)(1+o_p(1))$, where
\benrr
 I_6^*&&=\frac{2}{Nn(n-1)}\sti\sum_{j\neq i}^n
\frac{K_h(z_i-z_j)e_i}{\bar f_{N(2)}(\b^{\top}w_j)}
\sum_{s=N/2+1}^{N}M_{v_N}(\b^{\top}w_j-\b^{\top}\ti
w_s)g{'}(\b^{\top}\ti x_s)\ti x_s^{\top}\\
&&=: \frac{1}{n(n-1)}\sti\sum_{i\neq j}^n K_h(z_i-z_j)e_i I_{62}(\b^{\top}w_j), \qquad \mb{say}.
\eenrr
It is easy to see that for any given $\b^{\top}w_j$, $E[I_{62}(\b^{\top}w_j)]=E[g{'}(\b^{\top}x)x^{\top}|\b^{\top}w_j]$ by noticing that $\ti x$ has the same distribution as that of $x$. By Lemma 2 of Guo et al. (2015),
\benrr
\frac{1}{n(n-1)}\sti\sum_{i\neq j}^n K_h(z_i-z_j)e_iE[g{'}(\b^{\top}x)x^{\top}|\b^{\top}w_j]=O_p(\frac{1}{\sqrt{n}}).
\eenrr
Similarly, as in the proof for $I_4$, we can also derive that  as $N\rightarrow\infty$, $\sup_{\b^{\top}w}|I_{62}(\b^{\top}w)-E[I_{62}(\b^{\top}w)]|\leq O(v_N^2+\log(N)/\sqrt{Nv_N})$ and then
\benrr
\frac{1}{n(n-1)}\sti\sum_{i\neq j}^n K_h(z_i-z_j)e_i(I_{62}(\b^{\top}w_j)-E[g{'}(\b^{\top}x)x^{\top}|\b^{\top}w_j])=o_p(\frac{1}{\sqrt{n}}).
\eenrr
Hence $nh^{1/2}I_6=o_p(1)$.

Combining the above results for $I_4$, $I_5$ and $I_6$ with the fact that the remaining $21$ terms tend to zero, in probability, we  obtain that $nh^{1/2}V_{n2,1}\to_D N(0, 2\la^{-1}\t_2)$, where $\t_2$ is in (\ref{si123}).
\vs .1cm
{\it Step 5.2.2.} Next, consider the second term $V_{n2,2}$ of the decomposition (\ref{V12}). Rewrite
\benrr
V_{n2,2}&=&\frac{1}{n(n-1)}\sum_{i=1}^{n}\sum_{j\neq i}^{n}K_{h}(z_i-z_j)G_i(r_{j}-\h r_{j(2)})\\
        & &+\frac{1}{n(n-1)}\sum_{i=1}^{n}\sum_{j\neq i}^{n}(K_{h}(\h z_i-\h z_j)-K_{h}(z_i-z_j))G_i(r_{j}-\h r_{j(2)}).
\eenrr
Similarly as  the decomposition in (\ref{decomr}), $V_{n2,2}$ can also be decomposed
into 24 terms. Again, we only give the detail about how to treat the three leading
terms. Again, the remaining 21 terms tend to zero, in probability.
 The three leading terms are:
\benr
I_7 &= &\frac{1}{n(n-1)}\sum_{i=1}^{n}\sum_{j\neq i}^{n}K_{h}(z_i-z_j)G_i Q_{1(2)}(\b^{\top}w_j)/\bar f_{N(2)}(\b^{\top}w_j),\cr
I_8 &= &\frac{1}{n(n-1)}\sum_{i=1}^{n}\sum_{j\neq i}^{n}K_{h}(z_i-z_j)G_i Q_{2(2)}(\b^{\top}w_j)/\bar f_{N(2)}(\b^{\top}w_j),\nn \cr
I_9 &= &\frac{1}{n(n-1)}\sum_{i=1}^{n}\sum_{j\neq i}^{n}K_{h}(z_i-z_j)G_i Q_{3(2)}(\b^{\top}w_j)/\bar f_{N(2)}(\b^{\top}w_j),\nn
\eenr
where $Q_{1(2)}(\b^{\top}w_j)$, $Q_{2(2)}(\b^{\top}w_j)$, $Q_{3(2)}(\b^{\top}w_j)$ and $\bar f_{N(2)}(\b^{\top}w_j)$ are defined in (\ref{Q123(2)}) and (\ref{deffN2}).
Recall that $C_n=n^{-1/2}h^{-1/4}$ and $E[Q^2_{1(2)}(\b^{\top}w_j)]\leq C_1 v_N^{2\ell}+C_2v_N/N$, which was  proved when we handled $I_4$.
By the Cauchy--Schwarz inequality,
$$
|nh^{1/2}C_nI_{7}|\leq O_p\Big(n^{1/2}h^{1/4}\sqrt{C_1 v_N^{2\ell}+C_2v_N/N}\,\Big)=o_p(1).
$$

To deal with $I_8$, decompose $I_8=I_{81}+I_{82}$,  with
\benrr
&&I_{81}=\frac{1}{n(n-1)}\sum_{i=1}^{n}\sum_{j\neq i}^{n}K_{h}(z_i-z_j)G_i Q_{2(2)}(\b^{\top}w_j)/ \bar f(\b^{\top}w_j),\cr
&&I_{82}=\frac{1}{n(n-1)}\sum_{i=1}^{n}\sum_{j\neq i}^{n}K_{h}(z_i-z_j)G_i Q_{2(2)}(\b^{\top}w_j)[\frac{1}{\bar f_{N(2)}(\b^{\top}w_j)}-\frac{1}{\bar f(\b^{\top}w_j)}],
\eenrr
where $\bar f(\b^{\top}w)$ is the density of $\b^{\top}w$.
By some elementary calculations, one can verify that
$E[I_{81}^2]=O_p(1/N)$.
This implies $nh^{1/2}C_n I_{81}=o_p(1)$ by recalling the definition of $C_n$.

Next, consider $I_{82}$. By the Cauchy--Schwarz inequality, $I^2_{82}$ is
bounded above by a product of $\stj I^2_{821}(z_j)/n$ and
$\stj I^2_{822}(w_j)/n$, where
\benrr
I_{821}(z_j)=\frac{1}{n}\sum_{i\neq j}K_{h}(z_i-z_j)G_i,\quad I_{822}(w_j)=Q_{2(2)}(\b^{\top}w_j)[\frac{1}{\bar f_{N(2)}(\b^{\top}w_j)}-\frac{1}{\bar f(\b^{\top}w_j)}].
\eenrr

Now  we  bound $E[I^2_{821}(z_j)]$ and $E[I^2_{822}(w_j)]$.
 Clearly, conditional on $z_j$, $E[I^2_{821}(z_j)]=O(1)$, which in turn implies that
$E\big \{\stj I^2_{821}(z_j)/n\big \}=O(1).$

Next, note that

\benrr \frac 1n\stj I^2_{822}(w_j)&\le& \frac 1n\stj Q_{2(2)}^2(\b^{\top}w_j) \sup_{w}|\frac{1}{\bar f_{N(2)}(\b^{\top}w)}-\frac{1}{\bar f(\b^{\top}w)}|^2\\
&\le & O_p(v_N^{2}+\log(N)/\sqrt{Nv_N})\frac 1n\stj Q_{2(2)}^2(\b^{\top}w_j).
\eenrr
The second inequality is from the fact that   $\bar f(\b^{\top}w)$ is bounded below and $\sup_{w}|\bar f_{N(2)}(\b^{\top}w)-\bar f(\b^{\top}w)|=O_p(v_N^{2}+\log(N)/\sqrt{Nv_N})$.
By $E[(\ti r_s-\ti g_s)|\b^{\top}\ti w_s]=0$,  $E[Q^2_{2(2)}(\b^{\top}w_j)]\leq O(1/(Nv_N))$ for any fixed $\b^{\top}w_j$. In other words, $E\big \{\stj Q_{2(2)}^2(\b^{\top}w_j)/n\big \}\leq O(1/(Nv_N)).$  By the Markov inequality, $\stj I^2_{822}(w_j)/n$ is bounded by $O_p(1/{Nv_N})O_p({v_N^2+\log(N)/\sqrt{Nv_N}})=o_p(1/(nh^{1/2}C_n)^2).$ Combining these results, we obtain that
\benrr
\big|nh^{1/2}C_nI_{82}\big|\le
nh^{1/2}C_n o_p(1/(nh^{1/2}C_n)) =o_p(1).
\eenrr
The above results about $I_{81}$ and $I_{82}$ in turn yield that $nh^{1/2}C_nI_8=o_p(1)$.
\vs .1cm

Now we analyze $I_9$. Recall the definitions that $G_i=G(B^{\top}x_i)$ and $\D_i=E[G(B^{\top}X)|Z=z_i]$. Write $I_{9}=I_{91}+I_{92}$, where
\benrr
&&I_{91}=\frac{1}{n(n-1)}\sum_{i=1}^{n}\sum_{j\neq i}^{n}K_{h}(z_i-z_j)\D_{i}Q_{3(2)}(\b^{\top}w_j)/\bar f_{N(2)}(\b^{\top}w_j)\cr
&&I_{92}=\frac{1}{n(n-1)}\sum_{i=1}^{n}\sum_{j\neq i}^{n}K_{h}(z_i-z_j)(G_i-\D_i)Q_{3(2)}(\b^{\top}w_j)/\bar f_{N(2)}(\b^{\top}w_j).
\eenrr
For $I_{92}$, $E[G_i-\D_i|Z_i]=0$. Thus, $nh^{1/2}I_{92}=o_p(1)$, at the same rate as $I_6$.  So $nh^{1/2}C_n I_{92}=o_p(1)$.

Next, we deal with $I_{91}$. Similar to $I_8$, rewrite $I_{91}=I_{911}+I_{912}$,  where
\benrr
&&I_{911}=\frac{1}{n(n-1)}\sum_{i=1}^{n}\sum_{j\neq i}^{n}K_{h}(z_i-z_j)\D_{i}Q_{3(2)}(\b^{\top}w_j)/\bar f(\b^{\top}w_j), \\
&&I_{912}=\frac{1}{n(n-1)}\sum_{i=1}^{n}\sum_{j\neq i}^{n}K_{h}(z_i-z_j)\D_{i}Q_{3(2)}(\b^{\top}w_j)[\frac{1}{\bar f_{N(2)}(\b^{\top}w_j)}-\frac{1}{\bar f(\b^{\top}w_j)}].
\eenrr
Similar to the proof of $I_{82}$,  we have  $nh^{1/2}I_{912}=o_p(1)$, because $E[Q^2_{3(2)}(\b^{\top}w_j)]= O_p(C_n^2)$.

Next, consider $I_{911}$.
Define
\benrr
I^{*}_{911}:=E[I_{911}|z_i, \ti z_s, \ti x_s]=\frac{2}{nN}\sum_{i=1}^{n}\sum_{s=N/2+1}^{N}K_{h}(z_i-\ti z_s)\D_i(\ti g_s-\h{\ti g}_s).
\eenrr
By the first order Taylor expansion,
\benrr
I^{*}_{911}=\frac{2}{nN}\sum_{i=1}^{n}\sum_{s=N/2+1}^{N}K_{h}(z_i-\ti z_s)\D_{i}g{'}(\b_0^{\top}\ti x_{s})\ti x_s^{\top}(\b_0-\h \b)(1+o_{p}(1))
\eenrr
Combining the result of (\ref{localbeta}),
\benrr
nh^{1/2}C_{n}I^{*}_{911}\to_p  &\nu_2=-E\{\D(Z)E[g{'}(\b_0^{\top}X)X^{\top}|Z]f(Z)\}H(\b_0).
\eenrr
By computing the second moment of $I_{911}-I^{*}_{911}$ and
using the Markov
inequality, one can verify
$nh^{1/2}C_n(I_{911}-I^{*}_{911})=o_p(1)$.
Hence $nh^{1/2}C_n I_9 \to \nu_2$.
These results about $I_7$, $I_8$ and $I_9$ imply that  $nh^{1/2}C_{n}V_{n2,2}\to_p  \nu_2$.
Hence Step~\ref{step5.2} is finished. \hfill $\fbox{}$

\vs .1cm
\begin{step}\label{step5.3} \,\,\,
$ nh^{1/2}V_{n3}\to_DN\left( \nu_2, 2\la^{-1}\t_2\right), $
where $\nu_2$ and $\t_2$ are as in (\r{nu2}) and (\r{si123}).
\end{step}
{\it Proof:}
The proof is similar to that pertaining to $V_{n2}$ in STEP \r{step5.2}. The only difference is that instead of the representation (\r{decomr}) we now use
\benr\label{decomr2}
r_{i}-\h r_{i(1)} =\frac{2}{N}\sum_{t=1}^{N/2}M_{v_N}(\h \b^{\top}w_i-\h \b^{\top}\ti w_t)(r_{i}-\h{\ti g}_{t})/\frac{2}{N} \sum_{t=1}^{N/2} M_{v_N}(\h \b^{\top}w_i-\h \b^{\top}\ti w_t).
\eenr
Further the definitions in (\ref{deffN2}) and (\ref{Q123(2)}) are changed into
\benr\label{deffN1}
\bar f_{N(1)}(x)=\frac{2}{N}\sum_{t=1}^{N/2}M_{v_N}(x-\b^{\top}\ti w_t),\,\, \h{\bar f}_{N(1)}(x) = \frac{2}{N} \sum_{t=1}^{N/2} M_{v_N}(x-\h \b^{\top}\ti w_t),
\eenr
and
\benr\lel{Q123(1)}
&&Q_{1(1)}(\b^{\top}w_i)=\frac{2}{N}\sum_{t=1}^{N/2}M_{v_N}(\b^{\top}w_i-\b^{\top}\ti w_t)(r_i-\ti r_t),\\
&&Q_{2(1)}(\b^{\top}w_i)=\frac{2}{N}\sum_{t=1}^{N/2}M_{v_N}(\b^{\top}w_i-\b^{\top}\ti w_t)(\ti r_t-\ti g_t),\nn \\
&&Q_{3(1)}(\b^{\top}w_i)=\frac{2}{N}\sum_{t=1}^{N/2}M_{v_N}(\b^{\top}w_i-\b^{\top}\ti w_t)(\ti g_t-\h{\ti g}_t). \nn
\eenr
We omit the details here. \hfill $\fbox{}$

\begin{step}\label{step5.4} \,\,
$nh^{1/2}V_{n4}\to_DN(\nu_3, 2\la^{-2}\t_3)$, where $\t_3$ is as in (\r{si123}) and
\benr\lel{nu3}
 \nu_3=H^{\top}(\b_0)E\{E[g{'}(\b_0^{\top}X)X|Z]E[g{'}(\b_0^{\top}X)^{\top}X^{\top}|Z]f(Z)\}H(\b_0).
\eenr
\end{step}
{\it Proof:}
By the same decompositions in (\ref{decomr}) and (\ref{decomr2}), $V_{n4}$ can be decomposed to 9 dominant terms, and seven of those are of order $o_{p}(1/nh^{1/2})$. We investigate the other two terms as follows:
\benrr
I_{10}&=&\frac{1}{n(n-1)}\sum_{i=1}^{n}\sum_{j\neq i}^{n}K_{h}(z_i-z_j)Q_{2(1)}(\b^{\top}w_i)Q_{2(2)}(\b^{\top}w_j)/\bar f_{N(1)}(\b^{\top}w_i)\bar f_{N(2)}(\b^{\top}w_j), \\
I_{11}&=&\frac{1}{n(n-1)}\sum_{i=1}^{n}\sum_{j\neq i}^{n}K_{h}(z_i-z_j)Q_{3(1)}(\b^{\top}w_i)Q_{3(2)}(\b^{\top}w_j)/\bar f_{N(1)}(\b^{\top}w_i)\bar f_{N(2)}(\b^{\top}w_j).
\eenrr
Similar to the proof of $I_5$, we have $Nh^{1/2}I_{10}\to_DN(0,2\t_3)$, where $\t_3$ is defined in (\ref{si123}). Similarly as $I_{91}$, $I_{11}$ can be rewritten as
\benrr
I_{11}&=&\frac{4}{N^2}\sum_{t=1}^{N/2}\sum_{s=N/2+1}^{N}K_{h}(\ti z_t-\ti z_s)(\ti g_s-\h{\ti g}_s)(\ti g_t-\h{\ti g}_t)(1+o_{p}(1))\\
     &=&(\b_0-\h \b)^{\top}\left[\frac{4}{N^2}\sum_{s=1}^{N/2}\sum_{t=N/2+1}^{N}K_{h}(\ti z_t-\ti z_s)g{'}(\b_0^{\top}\ti x_s)g{'}(\b_0^{\top}\ti x_t)\ti x_s \ti x_t^{\top}\right](\b_{0}-\h \b).
\eenrr
Combining the result of (\ref{localbeta}), $nh^{1/2}I_{11}$ converges to $\nu_3$ in probability.
Hence Step~\ref{step5.4} is completed. \hfill $\fbox{}$
\vs .1cm
Altogether, Steps~\ref{step5.1}-- \ref{step5.4} conclude the proof of (ii) in Theorem~\r{th3.5}.
\vs .1cm

Next, we give a sketch of the proof of  (i), which describes the asymptotic
power performance of the test under the global alternative with fixed $C_n\equiv C$. Let
\benrr
\ti \b=\arg\min_\b E\left\{Y-\bar W E^{-1}[\bar W\bar W^{\top}]E[\bar W g(\b^{\top}X)]\right\}^{2}
\eenrr
which is different from the true parameter $\b_0$.
Here $\bar W$ is a vector consisting  of polynomials of $W$.
Then, for fixed $C_n\equiv C$,
\benrr
\h e&=&e+C(G(B^{\top}W)-E[G(B^{\top}W)|\ti \b^{\top}W])+CE[G(B^{\top}W)|\ti \b^{\top}W]\\
            &&+(E[g(\b_0^{\top}X)|\ti \b^{\top}W]-E[g(\ti \b^{\top}X)|\ti \b^{\top}W])+(E[g(\ti \b^{\top}X)|\ti \b^{\top}W]-E[g(\h \b^{\top}X)|\h \b^{\top}W]).
\eenrr
We can obtain that $V_n$ tends, in probability, to a positive constant since the third term in the right hand side of the above equation is not 0.
Similarly, we can also prove that $\h \t$ converges to a positive constant. We then have that $V_{n}/\h \t$ converges in probability to a positive constant. That is, the test statistic $nh^{1/2} V_n$ goes to infinity at the rate of order $nh^{1/2}$. The proof is finished. \hfill $\fbox{}$

\subsection{Proof of Theorem~\ref{theo3.1}}
As the arguments used for proving Theorem~\ref{th3.5} with $C_n=0$, the results $\|\h B-B\|=O_p(1/\sqrt{n})$ and $\h \b-\b=O_p(1/\sqrt{n})$ are applicable for proving this theorem, we then omit  most of the details, but focus on the bias term.
The terms $\bar f_{N(j)}(x)$, $Q_{k(j)}(\cdot)$, $k=1,2,3$ and $j=1,2$ in the proof of Theorem~\ref{th3.5} are replaced by
\benr\label{deffN}
\bar f_{N}(x)=\frac{1}{N}\sum_{s=1}^{N}M_{v_N}(x-\b^{\top}\ti w_s),\,\, \h{\bar f}_{N}(x) = \frac{1}{N} \sum_{s=1}^{N} M_{v_N}(x-\h \b^{\top}\ti w_s)
\eenr
and
\benr\lel{Q123}
&&Q_{1}(\b^{\top}w_i)=\frac{1}{N}\sum_{s=1}^{N}M_{v_N}(\b^{\top}w_i-\b^{\top}\ti w_s)(r_i-\ti r_s),\\
&&Q_{2}(\b^{\top}w_i)=\frac{1}{N}\sum_{s=1}^{N}M_{v_N}(\b^{\top}w_i-\b^{\top}\ti w_s)(\ti r_s-\ti g_s),\nn \\
&&Q_{3}(\b^{\top}w_i)=\frac{1}{N}\sum_{s=1}^{N}M_{v_N}(\b^{\top}w_i-\b^{\top}\ti w_s)(\ti g_s-\h{\ti g}_s). \nn
\eenr
Using the same decomposition as in the proof of Step~\ref{step5.4}, we also have a term similar to $I_{10}$ with the conditional expectation as
\benrr
I_{10}=\frac{1}{n(n-1)}\sum_{i=1}^{n}\sum_{j\neq i}^{n}K_{h}(z_i-z_j)Q_{2}(\b^{\top}w_i)Q_{2}(\b^{\top}w_j)/\bar f_{N}(\b^{\top}w_i)\bar f_{N}(\b^{\top}w_j)
\eenrr
and
\benrr
E[I_{10}|\ti \eta_s, \ti z_s, \ti \eta_t, \ti z_t]&=&\frac{1}{N^{2}}\sum_{s=1}^{N}\sum_{t=1}^{N}
\frac{1}{h}K(\frac{\ti z_s-\ti z_t}{h})\ti \eta_s \ti \eta_t (1+o_p(1)).
\eenrr
Separate the summands with $s\neq t$ and $s=t$ to write the leading term in the above expression as the sum of the following two  terms.
\benrr
I_{101}^{*}=\frac{1}{N^{2}}\sum_{s=1}^{N}\sum_{t\neq s}^{N}\frac{1}{h}K(\frac{\ti z_s-\ti z_t}{h})\ti \eta_s\ti \eta_t, \quad
I_{102}^{*}=\frac{1}{N^{2}}\sum_{s=1}^{N}\frac{1}{h}K(0)\ti \eta_{s}^2.
\eenrr
Since $K$ is symmetric, $I_{101}^{*}$ can be written as an U-statistic  with the kernel
\benrr
H_n((\ti z_s, \ti \eta_s),(\ti z_t,\ti \eta_t))=\frac{1}{h}K(\frac{\ti z_s-\ti z_t }{h})\ti \eta_s \ti \eta_t.
\eenrr
Further,
\benrr
E[H_n((\ti z_s, \ti \eta_s),(\ti z_t,\ti \eta_t))|(\ti z_s, \ti \eta_s)]=\frac{1}{h}\ti \eta_s E\{K(\frac{\ti z_s-\ti z_t}{h})\times E[\ti \eta_t|\ti z_t]\}=0.
\eenrr
Thus the U-statistic $I_{101}^{*}$ is degenerate.
By Central Limit Theorem for degenerate U-statistic (see, Hall 1984),
\benrr
Nh^{1/2}I_{101}^{*}\to_DN(0,2\int K^{2}(u)du\int(\xi^{2}(z))^2f^{2}(z)dz).
\eenrr
Hence $nh^{1/2}I_{101}^{*}\to_DN(0,\la^{-2}\t_3)$, where $\t_3$ is defined in
(\ref{si123}).
 Further, the fact that $Nh EI_{102}^{*}= K(0)E[\xi^2(Z)]$  implies that  $nh^{1/2}EI_{102}^{*}\to \iny$, which results in the asymptotic bias in $\ti V_n$. \hfill $\fbox{}$

\subsection{Proof of Theorem~\ref{theo3.3}}

When $N/n\to  0$, $\h \b$ and $\h B$ are $\sqrt{N}$ consistent estimates of $\b$ and
$B$, respectively. Again as the decompositions used in the proof of
Theorem~\ref{th3.5} are applicable for proving this theorem, we give only a
sketch of the proof of (i) here. Put $C_n=0$ in the proof of Theorem~\ref{th3.5}.
We only consider $I_1$, $V_{n2,1}$, and $I_{10}$. As $(N v_N^{1/2})/(nh^{1/2})
\to  0$, $Nv_N^{1/2}I_{1,1}$ in Step~\ref{step5.1} is $o_p(1)$. In addition,
$Nh^2\to  \iny$ leads  to $Nv_N^{1/2}I_{1,2}=o_p(1)$. Thus $Nv_N^{1/2}I_{1}=o_p(1)$. For $V_{n2,1}$, following
the proof of Step~\ref{step5.2}, we obtain that $Nv_N^{1/2}I_4=o_p(1)$,
$Nv_N^{1/2}I_5=o_p(1)$, $Nv_N^{1/2}I_6=o_p(1)$. These imply that
$Nv_N^{1/2}V_{n2}=o_p(1)$. Recalling the notation in (\ref{defs1}), (\ref{defs2}), (\ref{deffN}) and (\ref{Q123}), $I_{10}$ can be written as
\benrr
I_{10}=\frac{1}{n(n-1)N^2}\sum_{i=1}^{n}\sum_{j\neq i}^{n}K_{h}(z_i-z_j)Q_{2}(\b^{\top}w_i)Q_{2}(\b^{\top}w_j)/\bar f_{N}(\b^{\top}w_i)\bar f_{N}(\b^{\top}w_j).
\eenrr
Again define its conditional expectation as
\benrr
I_{10}^{*}&=&E[I_{10}|\ti z_s,\ti \eta_s,\ti z_t,\ti \eta_t]\\
&=&\frac{1}{N^{2}}\sum_{s=1}^{N}\sum_{t=1}^{N}\ti \eta_s \ti \eta_{t} \int\int\frac{1}{h}K(\frac{z_i-z_j}{h})\frac{1}{v_N}M(\frac{\b^{\top} w_i-\b^{\top}\ti w_s}{v_N})\\
& & \times\frac{1}{v_N}M(\frac{\b^{\top} w_j-\b^{\top}\ti w_t}{v_N})d(\b^{\top}w_i) d(\b^{\top}w_j).
\eenrr
Note that $\b^{\top} w=\|\b\|z$. Thus,
\begin{equation*}
\begin{split}
&\int\int\frac{1}{h}K(\frac{z_i-z_j}{h})\frac{1}{v_N}M(\frac{\b^{\top} w_i-\b^{\top}\ti w_s}{v_N})\frac{1}{v_N}M(\frac{\b^{\top} w_j-\b^{\top} w_t}{v_N})d(\b^{\top} w_i) d(\b^{\top}w_j)\cr
=&\int\int\frac{1}{h}K(\frac{z_i-z_j}{h})\frac{\|\b\|}{v_N}M(\frac{z_i - \ti z_s}{v_N/\|\b\|})\frac{\|\b\|}{v_N}M(\frac{z_j-\ti z_t}{v_N/\|\b\|})d z_i d z_j\cr
=&\int\int \frac{1}{h}K(u)\frac{\|\b\|}{v_N}M(\frac{hu + z_j-\ti  z_{s}}{v_N/\|\b\|})\frac{\|\b\|}{v_N}M(\frac{ z_j- \ti z_t}{v_N/\|\b\|})d(z_j+uh) d z_j\cr
=&\int \frac{\|\b\|}{v_N}M(\frac{z_j-\ti z_s}{v_N/\|\b\|})\frac{\|\b\|}{v_N}M(\frac{z_j-\ti z_t}{v_N/\|\b\|}) d z_j\cr
&+\int \frac{\|\b\|}{v_N}M^{''}(\frac{z_j-\ti z_s}{v_N/\|\b\|})\frac{\|\b\|^2h^2}{v_N^2}\frac{\|\b\|}{v_N}M(\frac{z_j- \ti z_t}{v_N/\|\b\|}) d z_j.
\end{split}
\end{equation*}
Let $\ti v_N=v_N/\|\b\|$. Then we have
\begin{equation*}
\begin{split}
I_{10}^{*}=&\frac{1}{N^{2}}\sum_{s=1}^{N}\sum_{t=1}^{N} \ti\eta_s \ti \eta_t \int \frac{1}{\ti v_N}M(\frac{z_j-\ti z_s}{\ti v_N})\frac{1}{\ti v_N}M(\frac{z_j-\ti z_t}{\ti v_N}) d z_j\cr
=&\frac{1}{N^{2}}\sum_{s=1}^{N}\sum_{t\neq s}^{N} \ti\eta_s \ti\eta_t\int \frac{1}{\ti v_N}M(\frac{z_j-\ti z_s}{\ti v_N})\frac{1}{\ti v_N}M(\frac{z_j-\ti z_t}{\ti v_N}) d z_j\cr
&+\frac{1}{N^{2}}\sum_{s=1}^{N} \ti \eta_{s}^2 \int \frac{1}{\ti v_N}M(\frac{z_j-\ti z_s}{v_N})\frac{1}{\ti v_N}M(\frac{z_j-\ti z_s}{\ti v_N}) d z_j\cr
&+\frac{1}{N^{2}}\sum_{s=1}^{N}\sum_{t\neq s}^{N} \ti \eta_s \ti\eta_t\int \frac{1}{\ti v_N}M^{''}(\frac{z_j-\ti z_s}{\ti v_N})\frac{h^2}{\ti v_N^2}\frac{1}{\ti v_N}M(\frac{z_j- \ti z_t}{\ti v_N}) d z_j\cr
&+\frac{1}{N^{2}}\sum_{s=1}^{N} \ti \eta_s^2 \int \frac{1}{\ti v_N}M^{''}(\frac{z_j-\ti z_s}{\ti v_N})\frac{h^2}{\ti v_N^2}\frac{1}{\ti v_N}M(\frac{z_j- \ti z_s}{\ti v_N}) d z_j \cr
=&:I_{101}+I_{102}+I_{103}+I_{104}.
\end{split}
\end{equation*}
Rewrite $I_{101}$ as
\benrr
2\sum_{s=2}^{N}\sum_{t<s}^{N}\ti \eta_s \ti \eta_t \frac{1}{N^{2}}\int \frac{1}{\ti v_N}M(\frac{z_j-\ti z_s}{\ti v_N}) \frac{1}{\ti v_N}M(\frac{z_j-\ti z_t}{\ti v_N}) d z_j.
\eenrr
By Theorem~1 of Hall (1984), $Nv_{N}^{1/2}I_{101}\to_DN(0, \ti \t)$,  where
\benrr
\ti \t=2 \|\b\|\int \Big(\int M(u)M(u+v)du)^2 dv \int (\xi^2(z)\Big)^2 f^2(z) dz,
\quad \xi^2(z)=E[\eta^2|Z=z].
\eenrr
We also have in probability
\benrr
N\ti v_N I_{102}\to_p E[\int \frac{1}{\ti v_N}M(\frac{z_j-\ti z_s}{\ti v_N})M(\frac{z_j-\ti z_s}{\ti v_N}) d z_j \ti \eta_s^2]=\int M^2(u)du E[\xi^2( z)].
\eenrr
 Further it can be proved that
\benrr
E[I_{103}^2] &=&O_p(\frac{h^4}{\ti v_N^4}\frac{1}{N^2\ti v_N})=o_p(\frac{1}{N^2 v_N}), \cr
E[I_{104}^2] &=&O_p(\frac{h^4}{\ti v_N^4}\frac{1}{N^2\ti v_N^2})+O_p(\frac{h^4}{\ti v_N^4}\frac{1}{N^3\ti v_N^3})=o_p(\frac{1}{N^2v_N}).
\eenrr
Then the Markov inequality implies that both $I_{103}$ and $I_{104}$ converge in probability to zero at the faster rate than $1/(Nv_{N}^{1/2})$.
We have $Nv_{N}^{1/2}\{I_{10}^{*}-\nu\}\to_DN(0,\ti\t)$. We can further prove that
\benrr
E[(I_{10}-I_{10}^{*})^2]=O_p(\frac{1}{N^2nv_N})=o_p(\frac{1}{N^2v_N}).
\eenrr
Hence $Nv_{N}^{1/2}\{I_{10}-\nu\}\to_DN(0,\ti\t)$. This completes the proof of Theorem \r{theo3.3}.
\hfill $\fbox{}$

\newpage

\begin{figure}
\begin{center}
\includegraphics[width=12cm,height=12cm]{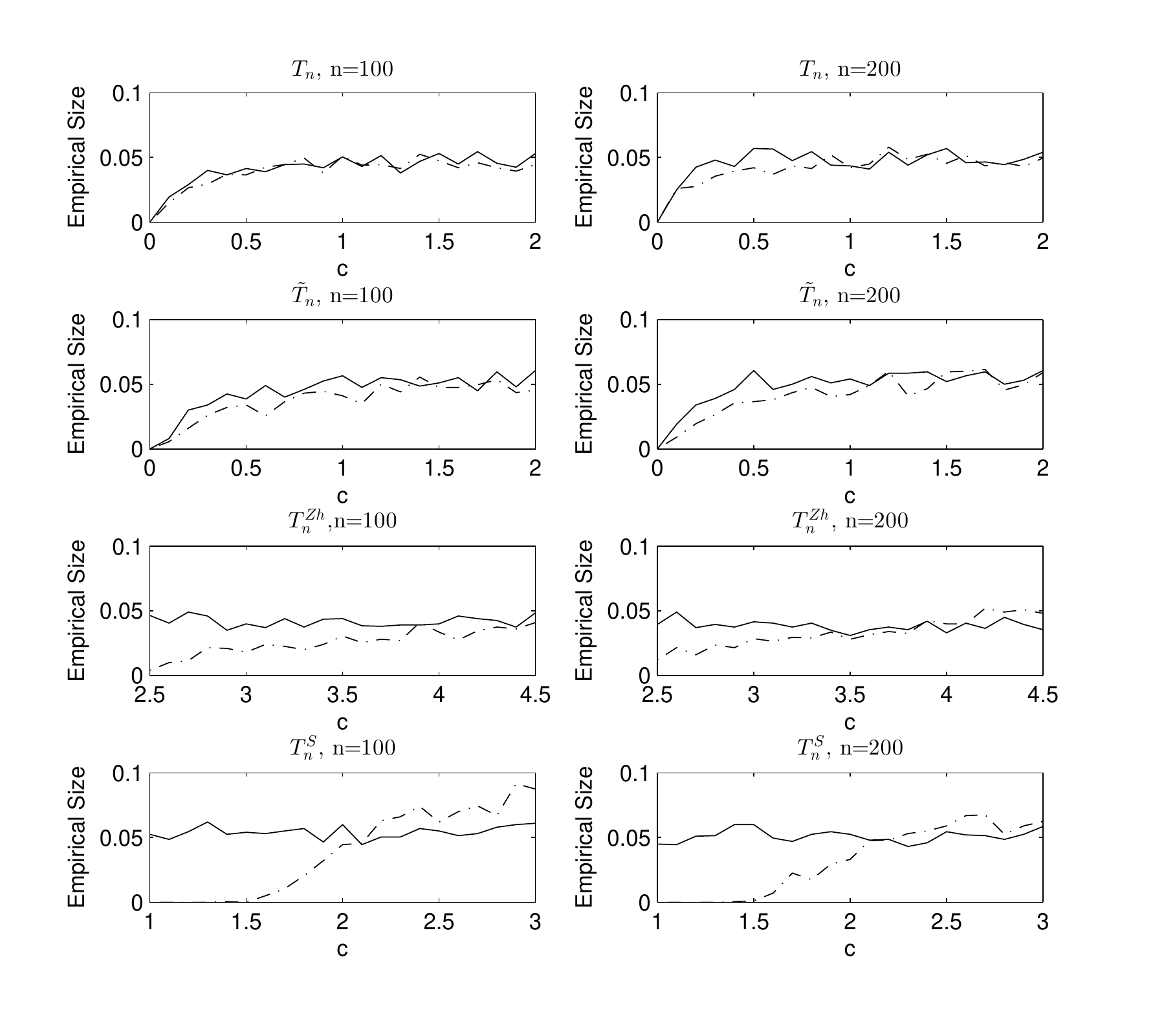}
\end{center}
\caption{Plots for the empirical size curve  against different values of $c$ in the bandwidths $h=cn^{-1/(4+q)}$, $v_N=c(N/2)^{-2/5}$. For model $Y=\b^{\top}X+\ep$, the solid lines are with $p=2$, $q=1$ and the dash-dotted lines are with $p=8$, $q=1$. }
\label{Fig1}
\end{figure}

\begin{table}
{Table~1. Empirical sizes and powers of $T_{n}$, ${T}_{n}^b$, $T_{n}^{Zh}$ and $T^{S}_{n}$ of $H_{0}$ vs. $H_{11}$ in Study~1.\vspace{0.3cm}}
\center {\small \scriptsize\hspace{12.5cm}
\renewcommand{\arraystretch}{1.2}\tabcolsep 0.2cm
\centering
\begin{tabular}{r|rrrrrrrrrr}
     \hline
    H11       & a     & \multicolumn{2}{c}{p=2} & \multicolumn{2}{c}{p=8} & \multicolumn{2}{c}{p=2} & \multicolumn{2}{c}{p=8} \\
    \hline
   $\la=4$&       & \multicolumn{2}{c}{$\Si=\Si_1$} & \multicolumn{2}{c}{$\Si=\Si_1$} & \multicolumn{2}{c}{$\Si=\Si_2$} & \multicolumn{2}{c}{$\Si=\Si_2$} \\
              &       & n=100  & n=200  & n=100  & n=200  & n=100  & n=200  & n=100  & n=200 \\
    \hline
 $T_{n}$      & 0     & 0.0455 & 0.0430 & 0.0420 & 0.0410 & 0.0495 & 0.0525 & 0.0505 & 0.0535 \\
              & 0.1   & 0.0700 & 0.0860 & 0.0715 & 0.0835 & 0.0720 & 0.1155 & 0.0825 & 0.1580 \\
              & 0.2   & 0.1275 & 0.2190 & 0.1185 & 0.2145 & 0.1970 & 0.4005 & 0.2720 & 0.6260 \\
              & 0.3   & 0.2360 & 0.4985 & 0.2185 & 0.4865 & 0.4245 & 0.7840 & 0.5630 & 0.9510 \\
              & 0.4   & 0.4265 & 0.8050 & 0.3940 & 0.7840 & 0.6695 & 0.9670 & 0.8180 & 0.9965 \\
              & 0.5   & 0.6315 & 0.9570 & 0.5670 & 0.9295 & 0.8385 & 0.9975 & 0.9305 & 1.0000 \\
     \hline
$\ti T_n$  & 0     & 0.0485 & 0.0520 & 0.0440 & 0.0525 & 0.0440 & 0.0510 & 0.0485 & 0.0460 \\
              & 0.1   & 0.0645 & 0.0760 & 0.0505 & 0.0865 & 0.0790 & 0.1300 & 0.1070 & 0.1615 \\
              & 0.2   & 0.1130 & 0.2335 & 0.1230 & 0.2210 & 0.2010 & 0.4135 & 0.2720 & 0.6240 \\
              & 0.3   & 0.2530 & 0.5205 & 0.2245 & 0.4975 & 0.4110 & 0.7900 & 0.5845 & 0.9500 \\
              & 0.4   & 0.4365 & 0.8055 & 0.3800 & 0.7980 & 0.6945 & 0.9720 & 0.8125 & 0.9930 \\
              & 0.5   & 0.6475 & 0.9495 & 0.5715 & 0.9360 & 0.8545 & 0.9995 & 0.9280 & 1.0000 \\
    \hline
$T_{n}^{Zh}$  & 0     & 0.0360 & 0.0335 & 0.0285 & 0.0410 & 0.0400 & 0.0385 & 0.0350 & 0.0405 \\
              & 0.1   & 0.0525 & 0.0940 & 0.0420 & 0.0525 & 0.0735 & 0.1060 & 0.0615 & 0.0925 \\
              & 0.2   & 0.1410 & 0.2475 & 0.0690 & 0.1045 & 0.2295 & 0.4280 & 0.1405 & 0.2710 \\
              & 0.3   & 0.3015 & 0.5780 & 0.1165 & 0.2230 & 0.4970 & 0.8385 & 0.2740 & 0.5715 \\
              & 0.4   & 0.5200 & 0.8395 & 0.1770 & 0.3740 & 0.7655 & 0.9800 & 0.4675 & 0.8270 \\
              & 0.5   & 0.7105 & 0.9690 & 0.2875 & 0.5500 & 0.9065 & 0.9985 & 0.6190 & 0.9420 \\
    \hline
$T^{S}_{n}$   & 0     & 0.0495 & 0.0570 & 0.0440 & 0.0340 & 0.0655 & 0.0595 & 0.0430 & 0.0425 \\
              & 0.1   & 0.1460 & 0.2060 & 0.0785 & 0.1125 & 0.2010 & 0.3020 & 0.1450 & 0.2250 \\
              & 0.2   & 0.3615 & 0.6110 & 0.2030 & 0.3400 & 0.4895 & 0.8150 & 0.4015 & 0.7160 \\
              & 0.3   & 0.6235 & 0.9145 & 0.3665 & 0.6625 & 0.8045 & 0.9860 & 0.7030 & 0.9650 \\
              & 0.4   & 0.8580 & 0.9870 & 0.5555 & 0.8820 & 0.9610 & 0.9990 & 0.8895 & 0.9975 \\
              & 0.5   & 0.9550 & 0.9999 & 0.7305 & 0.9705 & 0.9895 & 1.0000 & 0.9715 & 1.0000 \\
    \hline
\end{tabular} }
\end{table}

\begin{table}
{Table~2. Empirical sizes and powers of $T_{n}$, $\ti T_n$, $T_{n}^{Zh}$ and $T^{S}_{n}$ of $H_{0}$ vs. $H_{12}$ in Study~1. \vspace{0.3cm}}
\center {\small \scriptsize\hspace{12.5cm}
\renewcommand{\arraystretch}{1.2}\tabcolsep 0.2cm
\begin{tabular}{r|rrrrrrrrrr}
 \hline
    H12   & a     & \multicolumn{2}{c}{p=2} & \multicolumn{2}{c}{p=8} & \multicolumn{2}{c}{p=2} & \multicolumn{2}{c}{p=8} \\
 \hline
$\la=4$&       & \multicolumn{2}{c}{$\Si=\Si_{1}$} & \multicolumn{2}{c}{$\Si=\Si_{1}$} & \multicolumn{2}{c}{$\Si=\Si_{2}$} & \multicolumn{2}{c}{$\Si=\Si_{2}$} \\
          &       & n=100  & n=200  & n=100  & n=200  & n=100  & n=200 & n=100  & n=200 \\
    \hline
$T_{n}$   & 0     & 0.0480 & 0.0555 & 0.0410 & 0.0440 & 0.0525 & 0.0465 & 0.0475 & 0.0410 \\
          & 0.1   & 0.0520 & 0.1020 & 0.0595 & 0.0885 & 0.0625 & 0.0990 & 0.0495 & 0.0675\\
          & 0.2   & 0.1315 & 0.2350 & 0.1258 & 0.2140 & 0.1340 & 0.2080 & 0.1075 & 0.1835 \\
          & 0.3   & 0.2465 & 0.4935 & 0.2245 & 0.4545 & 0.2375 & 0.4580 & 0.1875 & 0.3755 \\
          & 0.4   & 0.4260 & 0.7585 & 0.3660 & 0.7250 & 0.3970 & 0.7020 & 0.2980 & 0.6045 \\
          & 0.5   & 0.6310 & 0.9220 & 0.5685 & 0.9105 & 0.5815 & 0.8840 & 0.4665 & 0.8155 \\
     \hline
$\ti T_n$& 0   & 0.0445 & 0.0490 & 0.0500 & 0.0515 & 0.0555 & 0.0480 & 0.0475 & 0.0410 \\
          & 0.1   & 0.0705 & 0.0825 & 0.0625 & 0.0790 & 0.0635 & 0.0855 & 0.0695 & 0.0820\\
          & 0.2   & 0.1375 & 0.2280 & 0.1130 & 0.2245 & 0.1425 & 0.2235 & 0.1055 & 0.1880 \\
          & 0.3   & 0.2805 & 0.4830 & 0.2280 & 0.4630 & 0.2545 & 0.4335 & 0.1995 & 0.3615 \\
          & 0.4   & 0.4415 & 0.7750 & 0.3700 & 0.7410 & 0.4165 & 0.7050 & 0.3120 & 0.6335 \\
          & 0.5   & 0.6315 & 0.9250 & 0.5875 & 0.9165 & 0.5705 & 0.8935 & 0.4650 & 0.8275 \\
    \hline
$T_{n}^{Zh}$& 0   & 0.0330 & 0.0425 & 0.0300 & 0.0400 & 0.0390 & 0.0495 & 0.0420 & 0.0405 \\
          & 0.1   & 0.0670 & 0.0995 & 0.0400 & 0.0500 & 0.0585 & 0.0930 & 0.0445 & 0.0640 \\
          & 0.2   & 0.1535 & 0.2520 & 0.0615 & 0.1065 & 0.1425 & 0.2340 & 0.0655 & 0.0975 \\
          & 0.3   & 0.3005 & 0.5330 & 0.1215 & 0.2320 & 0.2620 & 0.4795 & 0.0990 & 0.1845 \\
          & 0.4   & 0.5000 & 0.7975 & 0.2040 & 0.3825 & 0.4590 & 0.7525 & 0.1630 & 0.3225 \\
          & 0.5   & 0.7060 & 0.9445 & 0.3060 & 0.5900 & 0.6620 & 0.9115 & 0.2500 & 0.4865 \\
    \hline
$T^{S}_{n}$& 0    & 0.0530 & 0.0510 & 0.0460 & 0.0365 & 0.0505 & 0.0475 & 0.0450 & 0.0365 \\
          & 0.1   & 0.0100 & 0.1390 & 0.0715 & 0.0805 & 0.0855 & 0.1335 & 0.0580 & 0.0805 \\
          & 0.2   & 0.2135 & 0.3790 & 0.1470 & 0.2305 & 0.1985 & 0.3290 & 0.1240 & 0.1765 \\
          & 0.3   & 0.4385 & 0.6930 & 0.2625 & 0.4995 & 0.3695 & 0.6185 & 0.2005 & 0.3680 \\
          & 0.4   & 0.6710 & 0.9050 & 0.4420 & 0.7505 & 0.5720 & 0.8685 & 0.3130 & 0.5885 \\
          & 0.5   & 0.8375 & 0.9825 & 0.6265 & 0.9170 & 0.7670 & 0.9645 & 0.4890 & 0.8050 \\
\hline
\end{tabular} }
\end{table}

\begin{table}
{Table~3. Empirical sizes and powers of $T_{n}$, $\ti T_n$, $T_{n}^{Zh}$ and $T^{S}_{n}$ of $H_{0}$ vs. $H_{13}$ in Study~1. \vspace{0.3cm}}
\center {\small \scriptsize\hspace{12.5cm}
\renewcommand{\arraystretch}{1.2}\tabcolsep 0.2cm
\centering
\begin{tabular}{r|rrrrrrrrrr}
\hline
    H13   & a     & \multicolumn{2}{c}{p=2} & \multicolumn{2}{c}{p=8} & \multicolumn{2}{c}{p=2} & \multicolumn{2}{c}{p=8} \\
\hline
$\la=4$&       & \multicolumn{2}{c}{$\Si=\Si_{1}$} & \multicolumn{2}{c}{$\Si=\Si_{1}$} & \multicolumn{2}{c}{$\Si=\Si_{2}$} & \multicolumn{2}{c}{$\Si=\Si_{2}$} \\
          &       & n=100  & n=200  & n=100  & n=200  & n=100  & n=200  & n=100  & n=200 \\
\hline
$T_{n}$    & 0    & 0.0415 & 0.0505 & 0.0565 & 0.0455 & 0.0500 & 0.0420 & 0.0460 & 0.0495 \\
          & 0.1   & 0.0770 & 0.0900 & 0.0725 & 0.0860 & 0.0665 & 0.0735 & 0.0595 & 0.0705 \\
          & 0.2   & 0.1370 & 0.2470 & 0.1125 & 0.2115 & 0.1165 & 0.1885 & 0.0865 & 0.1550 \\
          & 0.3   & 0.2530 & 0.4430 & 0.2105 & 0.4130 & 0.2235 & 0.3920 & 0.1390 & 0.2980 \\
          & 0.4   & 0.3980 & 0.6965 & 0.3480 & 0.6470 & 0.3185 & 0.6220 & 0.1980 & 0.4410 \\
          & 0.5   & 0.5395 & 0.8715 & 0.4515 & 0.8205 & 0.4425 & 0.7815 & 0.2810 & 0.6075 \\
\hline
$\ti T_n$& 0   & 0.0455 & 0.0530 & 0.0585 & 0.0455 & 0.0475 & 0.0565 & 0.0500 & 0.0485 \\
          & 0.1   & 0.0605 & 0.0910 & 0.0665 & 0.0805 & 0.0765 & 0.0965 & 0.0590 & 0.0725 \\
          & 0.2   & 0.1360 & 0.2420 & 0.1100 & 0.2240 & 0.1100 & 0.1980 & 0.0880 & 0.1570 \\
          & 0.3   & 0.2680 & 0.4595 & 0.2090 & 0.4440 & 0.2120 & 0.4065 & 0.1335 & 0.2905 \\
          & 0.4   & 0.3750 & 0.6920 & 0.3365 & 0.6405 & 0.3375 & 0.6135 & 0.1910 & 0.4665 \\
          & 0.5   & 0.5520 & 0.8730 & 0.4400 & 0.8375 & 0.4605 & 0.7775 & 0.2685 & 0.5910 \\
\hline
$T_{n}^{Zh}$& 0   & 0.0350 & 0.0450 & 0.0250 & 0.0450 & 0.0365 & 0.0505 & 0.0355 & 0.0415 \\
          & 0.1   & 0.0560 & 0.0875 & 0.0350 & 0.0410 & 0.0510 & 0.0610 & 0.0365 & 0.0445 \\
          & 0.2   & 0.1130 & 0.2250 & 0.0525 & 0.0875 & 0.0985 & 0.1650 & 0.0400 & 0.0600 \\
          & 0.3   & 0.2215 & 0.4460 & 0.0795 & 0.1380 & 0.1705 & 0.3570 & 0.0580 & 0.0860 \\
          & 0.4   & 0.3700 & 0.6760 & 0.1135 & 0.2265 & 0.3120 & 0.5650 & 0.0665 & 0.1295 \\
          & 0.5   & 0.5075 & 0.8410 & 0.1610 & 0.3225 & 0.4010 & 0.7330 & 0.0780 & 0.1650 \\
\hline
$T^{S}_{n}$& 0    & 0.0570 & 0.0410 & 0.0405 & 0.0420 & 0.0560 & 0.0565 & 0.0440 & 0.0400 \\
          & 0.1   & 0.0560 & 0.0695 & 0.0505 & 0.0390 & 0.0500 & 0.0650 & 0.0555 & 0.0300 \\
          & 0.2   & 0.0945 & 0.1305 & 0.0750 & 0.0945 & 0.0640 & 0.0840 & 0.0610 & 0.0380 \\
          & 0.3   & 0.1455 & 0.2065 & 0.1150 & 0.1550 & 0.0870 & 0.0990 & 0.0520 & 0.0615 \\
          & 0.4   & 0.2030 & 0.3225 & 0.1550 & 0.2560 & 0.1120 & 0.1400 & 0.0625 & 0.0665 \\
          & 0.5   & 0.2540 & 0.4255 & 0.1895 & 0.3600 & 0.1350 & 0.1840 & 0.0660 & 0.0600 \\
\hline
\end{tabular} }
\end{table}

\begin{table}
{Table~4. Empirical sizes and powers of $T_{n}$ and $T^{S}_{n}$ of $H_{0}$ vs. $H_{14}$ and $H_{15}$ in Study~2. \vspace{0.3cm}}
\center {\small \scriptsize\hspace{12.5cm}
\renewcommand{\arraystretch}{1.2}\tabcolsep 0.2cm
\begin{tabular}{r|rrrrrrrrrr}
    \hline
          & a     &      \multicolumn{4}{c}{$H_{14}$} & \multicolumn{4}{c}{$H_{15}$}  \\
    \hline
$\la=4$&       & \multicolumn{2}{c}{$\Si=\Si_{1}$} & \multicolumn{2}{c}{$\Si=\Si_{2}$} & \multicolumn{2}{c}{$\Si=\Si_{1}$} & \multicolumn{2}{c}{$\Si=\Si_{2}$} \\
          &       & n=100  & n=200  & n=100  & n=200  & n=100  & n=200  & n=100  & n=200 \\
    \hline
$T_{n}$   & 0     & 0.0525 & 0.0470 & 0.0460 & 0.0485 & 0.0440 & 0.0450 & 0.0395 & 0.0460 \\
          & 0.1   & 0.0530 & 0.0720 & 0.0650 & 0.0805 & 0.0455 & 0.0430 & 0.0515 & 0.0710 \\
          & 0.2   & 0.0780 & 0.1245 & 0.1130 & 0.1720 & 0.0700 & 0.0700 & 0.1175 & 0.2020 \\
          & 0.3   & 0.1390 & 0.2385 & 0.1905 & 0.3865 & 0.0905 & 0.1455 & 0.1890 & 0.3920 \\
          & 0.4   & 0.2065 & 0.3660 & 0.2885 & 0.5860 & 0.1175 & 0.2490 & 0.2285 & 0.5200 \\
          & 0.5   & 0.3060 & 0.5560 & 0.4405 & 0.7890 & 0.1485 & 0.3130 & 0.2690 & 0.6105 \\
    \hline
$T^{S}_{n}$& 0    & 0.0525 & 0.0605 & 0.0605 & 0.0540 & 0.0450 & 0.0515 & 0.0540 & 0.0535 \\
          & 0.1   & 0.0830 & 0.0970 & 0.0915 & 0.1155 & 0.0620 & 0.0545 & 0.0525 & 0.0490 \\
          & 0.2   & 0.1375 & 0.2190 & 0.1755 & 0.3390 & 0.0575 & 0.0555 & 0.0450 & 0.0525 \\
          & 0.3   & 0.2310 & 0.4245 & 0.3575 & 0.6170 & 0.0485 & 0.0465 & 0.0590 & 0.0570 \\
          & 0.4   & 0.3615 & 0.6375 & 0.5205 & 0.8340 & 0.0530 & 0.0540 & 0.0550 & 0.0590 \\
          & 0.5   & 0.5020 & 0.8040 & 0.6935 & 0.9410 & 0.0590 & 0.0515 & 0.0505 & 0.0410 \\
     \hline
\end{tabular} }
\end{table}

\begin{table}
{Table~5. Empirical sizes and powers of $T_{n}$ and $T_{n}^{(1)}$(with small $\la$), $T_{n}^{(2)}$(with large $\la$) of $H_{0}$ vs. $H_{11}$ in Study~3. \vspace{0.3cm}}
\center {\small \scriptsize\hspace{12.5cm}
\renewcommand{\arraystretch}{1.2}\tabcolsep 0.2cm
\begin{tabular}{r|rrrrrrrrr}
\hline
$H_{11}$     &       & \multicolumn{2}{c}{p=2} & \multicolumn{2}{c}{p=8} & \multicolumn{2}{c}{p=2} & \multicolumn{2}{c}{p=8} \\
             &       & \multicolumn{2}{c}{$\la=0.1$} & \multicolumn{2}{c}{$\la=0.1$} & \multicolumn{2}{c}{$\la=0.5$} & \multicolumn{2}{c}{$\la=0.5$} \\
             &  a    & N=100  & N=200  & N=100  & N=200  & N=100  & N=200  & N=100  & N=200 \\
\hline
$T_{n}$      & 0     & 0.0160 & 0.0255 & 0.0080 & 0.0120 & 0.0330 & 0.0420 & 0.0235 & 0.0295 \\
             & 0.1   & 0.0380 & 0.0865 & 0.0280 & 0.0535 & 0.0535 & 0.0725 & 0.0425 & 0.0685\\
             & 0.2   & 0.1710 & 0.4420 & 0.1305 & 0.4305 & 0.1245 & 0.2400 & 0.0970 & 0.2265 \\
             & 0.3   & 0.4695 & 0.8920 & 0.4465 & 0.8835 & 0.2720 & 0.6005 & 0.2370 & 0.5905 \\
             & 0.4   & 0.7775 & 0.9935 & 0.7980 & 0.9930 & 0.4955 & 0.8990 & 0.4445 & 0.8765 \\
             & 0.5   & 0.9465 & 1.0000 & 0.9360 & 1.0000 & 0.7270 & 0.9860 & 0.6390 & 0.9805 \\
\hline
$T_{n}^{(1)}$& 0     & 0.0610 & 0.0555 & 0.0400 & 0.0475 & 0.1690 & 0.1720 & 0.1190 & 0.1490 \\
             & 0.1   & 0.1135 & 0.1745 & 0.0885 & 0.1635 & 0.2175 & 0.2470 & 0.1745 & 0.2600 \\
             & 0.2   & 0.3705 & 0.6415 & 0.3095 & 0.6200 & 0.3470 & 0.5370 & 0.3135 & 0.5295 \\
             & 0.3   & 0.7100 & 0.9680 & 0.6550 & 0.9595 & 0.5695 & 0.8410 & 0.5100 & 0.8165 \\
             & 0.4   & 0.9255 & 0.9995 & 0.9145 & 0.9995 & 0.7765 & 0.9715 & 0.7300 & 0.9605 \\
             & 0.5   & 0.9865 & 1.0000 & 0.9860 & 1.0000 & 0.9115 & 0.9975 & 0.8625 & 0.9985 \\
\hline
             &       & \multicolumn{2}{c}{$\la=4$} & \multicolumn{2}{c}{$\la=4$} & \multicolumn{2}{c}{$\la=8$} & \multicolumn{2}{c}{$\la=8$} \\
             &  a    & n=100  & n=200  & n=100  & n=200  & n=100  & n=200  & n=100  & n=200 \\
\hline
$T_{n}$      & 0     & 0.0525 & 0.0545 & 0.0480 & 0.0405 & 0.0485 & 0.0385 & 0.0430 & 0.0545 \\
             & 0.1   & 0.0590 & 0.0960 & 0.0530 & 0.0925 & 0.0705 & 0.0850 & 0.0615 & 0.0780 \\
             & 0.2   & 0.1270 & 0.2335 & 0.1110 & 0.2290 & 0.1325 & 0.2560 & 0.1340 & 0.2530 \\
             & 0.3   & 0.2645 & 0.5715 & 0.2525 & 0.5445 & 0.3045 & 0.5815 & 0.2550 & 0.5605 \\
             & 0.4   & 0.4390 & 0.8310 & 0.4175 & 0.8260 & 0.5030 & 0.8675 & 0.4445 & 0.8350 \\
             & 0.5   & 0.6705 & 0.9700 & 0.6295 & 0.9665 & 0.6885 & 0.9690 & 0.6620 & 0.9690 \\
\hline
$T_{n}^{(2)}$& 0     & 0.0610 & 0.0620 & 0.0575 & 0.0495 & 0.0530 & 0.0420 & 0.0445 & 0.0575 \\
             & 0.1   & 0.0660 & 0.1075 & 0.0685 & 0.1085 & 0.0755 & 0.0890 & 0.0690 & 0.0840 \\
             & 0.2   & 0.1410 & 0.2560 & 0.1310 & 0.2505 & 0.1450 & 0.2670 & 0.1430 & 0.2685 \\
             & 0.3   & 0.2910 & 0.5985 & 0.2845 & 0.5775 & 0.3145 & 0.5960 & 0.2735 & 0.5760 \\
             & 0.4   & 0.4720 & 0.8510 & 0.4490 & 0.8445 & 0.5175 & 0.8760 & 0.4620 & 0.8415 \\
             & 0.5   & 0.6880 & 0.9735 & 0.6580 & 0.9720 & 0.6950 & 0.9700 & 0.6745 & 0.9715 \\
\hline
\end{tabular} }
\end{table}

\newpage

\begin{figure}
\begin{center}
\includegraphics[width=12cm,height=10cm]{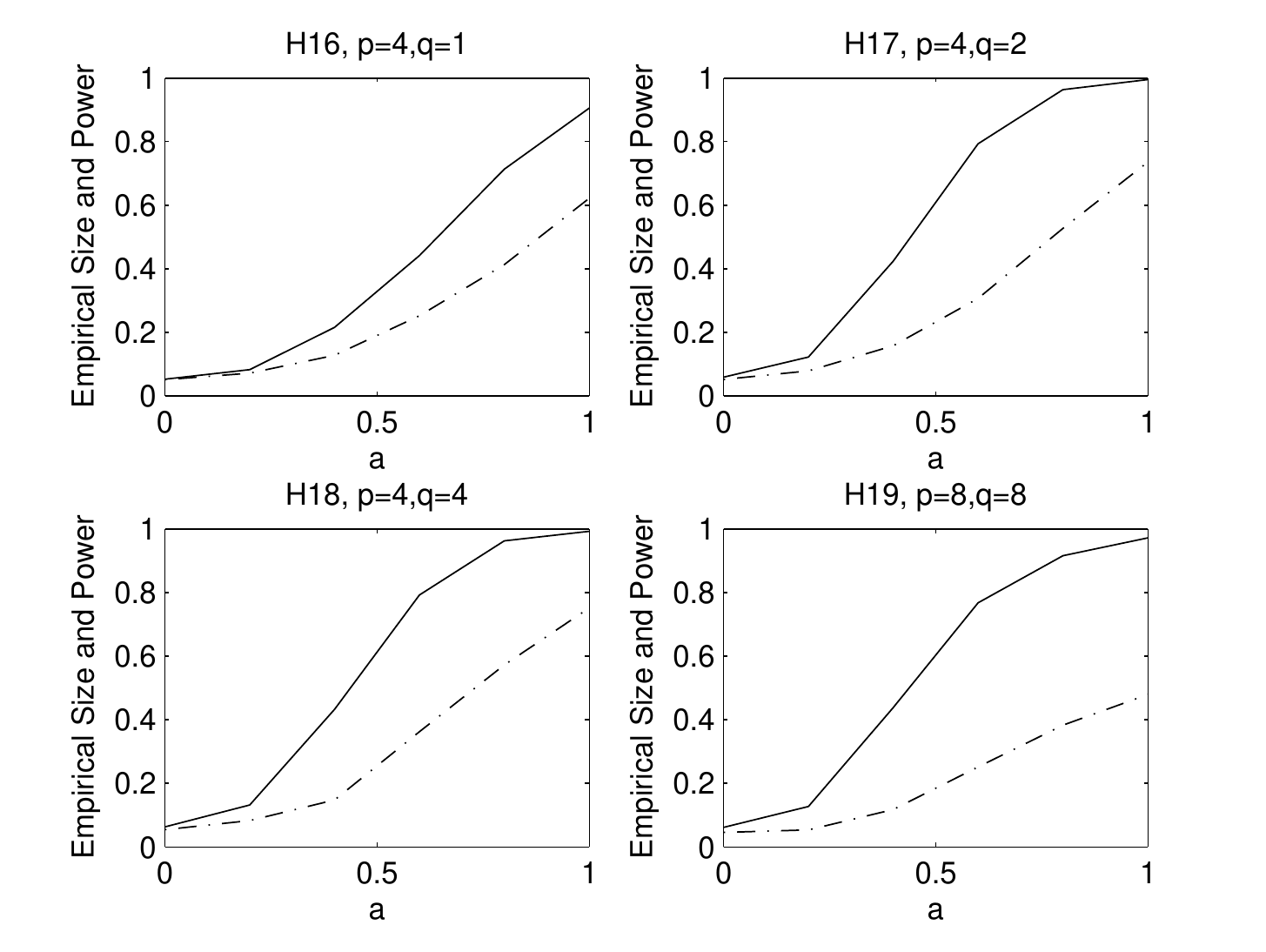}
\end{center}
\caption{Plots of power curves over $a$ under $H{16}- H{19}$ in Study~4. The solid lines are for  $T_n$  and the dash-dotted lines are for $T_n^{Zh}$. }
\label{Fig2}
\end{figure}

\end{document}